\documentclass{aa} 
\usepackage{amsmath,amssymb} 
\usepackage{graphicx}
\usepackage{natbib}
\bibpunct{(}{)}{;}{a}{}{,}
\usepackage{txfonts}
\usepackage[T1]{fontenc}
\usepackage{ulem}
\usepackage{xcolor}
\newcommand{\cmt}{\,cm$^{-3}$}   
   
\newcommand{\kms}{\,km\,s$^{-1}$}

\newcommand{\ecs}{$\rm\,erg\,cm^{-2}\,s^{-1}$}

\newcommand{\ro}{\,$R_{\odot}$}
\newcommand{\mo}{\,$M_{\odot}$}
\newcommand{\lo}{\,$L_{\odot}$}
\begin{document}

%
%
\title{Density asymmetry and wind velocities in the orbital plane of the symbiotic binary EG~Andromedae}   
\titlerunning{Density asymmetry and velocities of the wind in EG~And}
\authorrunning{N.~Shagatova, A.~Skopal, E.~Kundra et al.}

\author{N.~Shagatova    \inst{\ref{inst1}}
  \and A.~Skopal        \inst{\ref{inst1}}
  \and E.~Kundra        \inst{\ref{inst1}} 
  \and R.~Komžík        \inst{\ref{inst1}} 
  \and S.~Yu.~Shugarov  \inst{\ref{inst1}}
  \and T.~Pribulla      \inst{\ref{inst1}}  
  \and V.~Krushevska    \inst{\ref{inst1},\ref{inst3}}
  }
\institute{Astronomical Institute, Slovak Academy of Sciences,
           059~60 Tatransk\'{a} Lomnica, Slovakia
           \email{nshagatova@ta3.sk}\label{inst1} 
      \and Main Astronomical Observatory of National Academy of Sciences of Ukraine, 27 Akademika Zabolotnoho St., 031~43, Kyiv, Ukraine\label{inst3}
}

\date{Received / Accepted }

\abstract
{
Non-dusty late-type giants without a corona and large-scale pulsations represent objects that do not fulfil the conditions under which standard mass-loss mechanisms can be applied efficiently. Despite the progress during the past decades, the driving mechanism of their winds is still unknown.
}
{
One of the crucial constraints of aspiring wind-driving theories can be provided by the measured velocity and density fields of outflowing matter. The main goal of this work is to match the radial velocities of absorbing matter with a depth in the red giant (RG) atmosphere in the S-type symbiotic star \object{EG And}. 
}
{
We measured fluxes and radial velocities of ten Fe\,I absorption lines from spectroscopic observations with a resolution of $\approx 30\,000$. At selected orbital phases, we modelled their broadened profiles, including all significant broadening mechanisms.
}
{
The selected Fe\,I absorption lines at $5151$ - $6469$\,\AA\, originate at a radial distance $\approx 1.03$ RG radii from its centre. The corresponding radial velocity is typically $\approx 1$\kms\ , which represents a few percent of the terminal velocity of the RG wind. The high scatter of the radial velocities of several \kms\ in the narrow layer of the stellar atmosphere points to the complex nature of the near-surface wind mass flow.
The average rotational velocity of $11$\kms\
implies that the rotation of the donor star can contribute to observed focusing the wind towards the orbital plane.
The orbital variability of the absorbed flux indicates the highest column densities of the wind in the area between the binary components, even though the absorbing neutral material is geometrically more extended from the opposite side of the giant. This wind density asymmetry in the orbital plane region can be ascribed to gravitational focusing by the white dwarf companion.
}
{
Our results suggest that both gravitational and rotational focusing contribute to the observed enhancement of the RG wind towards the orbital plane, which makes mass transfer by the stellar wind highly efficient.
}
         
\maketitle

\section{Introduction}
\label{s:int}

The atmospheres of late-type giant stars include slow and dense winds reaching terminal velocities lower than $100$\kms with decreasing values for later spectral types \citep{du86}. For the asymptotic giant branch (AGB) evolutionary stage, the driving mechanism of the outflow is thought to be based on a combination of the dust-forming levitation of the wind by stellar pulsations and of the acceleration by radiation pressure on dusty envelopes \citep{ho+18}. On the other hand, the lack of dust in the atmospheres of normal red giant stars (RGs) and the inefficiency of other known driving mechanisms represent a complication for the understanding of their winds. Since the late 20th century, the dissipation of magnetic waves is thought to be the key ingredient in their mass-loss process. A review of attempts to resolve the mechanism behind RG  winds can be found in \cite{ha+80}, \cite{ho87}, \cite{ha96}, \cite{og+13}, and \cite{ai+15}. 
Recently, \cite{ha+22} investigated the wind properties of Arcturus (K1.5 III) using the Wentzel–Kramers–Brillouin Alfvén wave-driven wind
theory \citep{hm80}. They found that the wave periods that are required to match the observed damping rates correspond to hours to days, consistent with the photospheric granulation timescale.

The late-type giants play the role of donor star in the symbiotic stars (SySts), which are long-period ($P\gtrsim$ years) binary systems with a mass transfer of the giant wind towards a compact companion, usually a white dwarf \citep[e.g.][]{bo75,mu19}. The donor star supplying the dense wind matter is either a normal RG star (S-type SySts) or an AGB star (D-type SySts). The RGs in S-type systems have experienced the first dredge-up, which was confirmed by their low $^{12}$C/$^{13}$C ratio in the range $5$ - $23$ \citep{mi+14,ga+15,ga+16}.
The white dwarf as a source of ultraviolet radiation enables us to probe the cool wind from the giant at different directions. For example, the continuum depression around the Ly-$\alpha$ line as a function of the orbital phase has shown a very slow wind velocity up to $1-2$ RG radii, $R_{\rm g}$, above the donor surface and a steep increase to the terminal velocity afterwards in S-type SySts \citep[][ also Fig.~\ref{fig:vrlaw} here]{vo91,du+99,sh+16}. For D-type systems, this way of deriving the wind velocity profile is complicated by very long ($P\approx 10-100$\,yr) and often poorly known orbital periods. However, for single O-rich AGB stars, the expansion velocities of the wind were determined by \cite{ju+12} as a measure of the half-width of the molecular lines at the baseline level. The majority of the stars in their sample has a distinct low-velocity region in front of the velocity jump to the terminal value, but in a few cases, the wind reaches terminal velocity already within the innermost parts. In one case, the authors found a deceleration of the gas as it moves away from the star (R~Dor). The low-velocity region close to the star and a steep increase to the terminal velocity in O-rich AGB stars was also indicated by molecular line modelling with a non-local thermal equilibrium radiative transfer code \citep{de+06,de+10}. In C-rich AGB stars, the wind velocity profile can be steeper because the opacity of the dust grains is higher \citep[][ and references therein]{em+20}. For the C-rich AGB star CW Leo, a steep increase in the wind velocity was found to start at a distance of $\approx 5$ stellar radii \citep{de+15}.

The presence of the hot white dwarf \citep[$T_{\rm WD} \gtrsim 10^5$\,K;][]{mu+91,sk05} accompanied by the cool giant \citep[$T_{\rm RG} \approx 3000$ - $4000$\,K;][ and references therein]{ak+19} in S-type SySts leads to a complex ionization structure of the circumbinary material.
During quiescent phases when there is no ongoing eruptive burning on the surface of the white dwarf, a fraction of the surrounding RG wind is photoionized by energetic radiation from the hot component. As a result, the neutral area around the RG is cone-shaped, with the RG near its apex facing the white dwarf \citep{se+84,nu+87}, where a thin boundary between the neutral and ionized zone is determined by the balance between the flux of ionizing photons from the white dwarf and the flux of neutral particles from the RG.

\object{EG And} is an S-type SySt with no recorded outburst of its white dwarf. The effective temperature of the white dwarf is $\approx 7.5\times 10^4$\,K \citep{mu+91,vo+92} and its mass is $0.4\pm 0.1$\mo\, \citep{kg16,mi03}. The system is eclipsing \citep{vo91} with an orbital inclination of $\approx 80^\circ$ \citep{vo+92} and an orbital period of 483 days \citep{fe+00,kg16}.
The donor star is an RG of spectral class M2-3 III \citep{kf87,sc+92,mu+99} with an effective temperature $\approx 3700$\,K \citep[][ and references therein]{vo+92,kg16}, luminosity $\approx(1-2)\times 10^3$\lo\, \citep{vo+92,sk05}, and metallicity [Fe/H]$\approx 0$ \citep{wo+11}. Its mass is estimated to be $1.5\pm 0.6$\mo\, \citep{mi03} and its radius is estimated to be $75\pm 10$\,$R_\odot$ \citep{vo+92}, corresponding to $\log g \approx 0.5$ - $1.1$. The slow and dense wind of RG is assumed to have a terminal velocity $v_\infty \approx 30$\kms \citep{vo+92,lu+08}.
The velocity profile of the wind suggests an almost steady wind up to around 1.5\,R$_{\rm g}$ from the RG centre and subsequent rapid acceleration towards the terminal velocity, as derived from hydrogen column density values measured from the Ly$\alpha$-line attenuation \citep{sh+16}. This approach accounts for the wind density distribution at the near orbital plane due to the point-like relative size of the white dwarf as a source of the probing radiation.
The giant wind in this system is distributed asymmetrically, with denser parts concentrated at the orbital plane and diluted areas located around the poles \citep{sh+21}.

The geometric distribution and radial velocity (RV) profile of the RG wind are essential components for exploring the physical mechanism driving the outflow and shaping the RG wind. 
In this work, we analyse the orbital variability of fluxes and RVs of Fe\,I absorption lines of \object{EG And} (Sect.~\ref{ss:orbvar}). We intend to match the resulting RVs of individual lines with the depth of their origin in the atmosphere by modelling their profile using a semi-empirical model atmosphere (Sect.~\ref{ss:mod}) and including several broadening mechanisms (Sect.~\ref{ss:lprof}). The results are given in Sect.~\ref{ss:strat}. The discussion and conclusions can be found in Sects.~\ref{s:dis} and \ref{s:concl}, respectively.


\section{Observations}
\label{s:obs}

In the optical wavelength range, the main source of the continuum radiation in EG~And is the RG companion \citep{sk05}. Its spectrum is superposed with dominant Balmer emission lines arising in the symbiotic nebula and many absorption lines of molecules and atoms originating in the cool giant wind \citep[e.g.][]{kg16}. 

We collected 53 spectroscopic observations from Skalnaté Pleso Observatory (SP) from 2016 - 2023 in the wavelength range 4200 - 7300\,\r{A} (Table~\ref{tabAp:RVs} or \ref{tabAp:fluxes}). The observatory is equipped with a 1.3\,m Nasmyth-Cassegrain telescope (f/8.36) with a fibre-fed \'{e}chelle spectrograph (R$\sim$30\,000) similar to the MUSICOS design \citep[][]{bb92,pr+15}. 
The spectra were reduced with the Image Reduction and 
Analysis Facility (IRAF; \cite{to86}) using specific scripts and programs 
\citep[][]{pr+15}. The spectra were wavelength-calibrated using the ThAr hollow-cathode lamp. The achieved accuracy for our set of spectra corresponds to the systematic error of RV measurements, which typically is in the range 0.2 - 0.6\kms. 

Our spectra were dereddened with $E_{\rm B-V} = 0.05$\,mag 
\citep[][]{mu+91} using the extinction curve of \cite{ca+89}. 
We determined the orbital phase $\varphi$ of \object{EG And} using 
the ephemeris of the inferior conjunction of the RG 
($\varphi = 0$) given as \citep[][]{fe+00,kg16} 
\begin{equation}
\label{eq:ephem}
 JD_{\rm sp. conj.} = 
    2\,450\,683.2(\pm 2.4) + 482.6(\pm 0.5)\times E .
\end{equation}
We assumed a systemic velocity of 
$v_{\rm sys} = -94.88$\,kms$^{-1}$ \citep{kg16}. 
Similar values were determined by \cite{ol+85}, \cite{mu+88}, \cite{mu93}, and \cite{fe+00}.

We converted the spectra from relative into absolute fluxes by scaling them to the closest-date photometric fluxes using a fourth-degree polynomial function. We used the $UBVR_{\rm C}$ photometry of \object{EG And} published by \cite{se+19} together with new photometric observations obtained at the G2 pavilion of the Stará Lesná Observatory, which is equipped with a 60 cm, f/12.5 Cassegrain telescope \citep{se+19}. To complement our dataset during 2022, we used photometric observations available in the International Database of the American Association of Variable Star Observers (AAVSO\footnote{\url{https://aavso.org}}).
We converted the photometric magnitudes into fluxes according to the calibration in Table~2.2 of \cite{hk82}. 


\section{Analysis and results}
\label{anres}

To investigate the velocity distribution in the RG atmosphere of \object{EG And}, we selected ten Fe\,I absorption lines between 5151 and 6469\,\r{A} that were not severely blended.
We measured their orbital variability and modelled their absorption profiles to track the density conditions and dynamics of the corresponding part of the wind area.


\subsection{Orbital variations of the Fe\,I absorption lines}
\label{ss:orbvar}

The selected absorption lines of neutral iron show the orbital variability in RVs and absorbed fluxes. To measure these changes along the orbit, we fitted the lines with a Gaussian profile superimposed on a fourth-order polynomial function representing the continuum radiation of the spectrum (Sect.~\ref{s:obs}) using the curve-fitting program Fityk\footnote{\url{https://fityk.nieto.pl}} \citep{wo10}.

\begin{figure}
\centering
\begin{center}
\resizebox{\hsize}{!}{\includegraphics[angle=0]{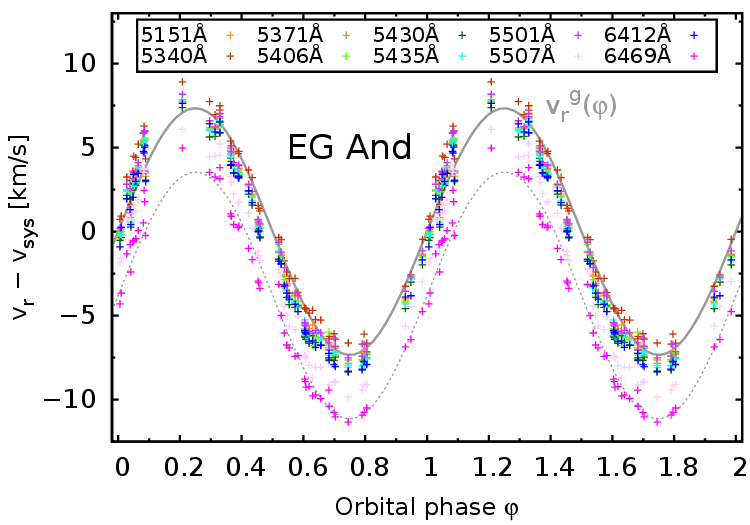}}
\end{center}
\caption{
Orbital variability in RVs of individual Fe\,I absorption lines (the corresponding colours are given in the legend). The solid grey line represents the RV curve of the RG, $v_r^{\rm g}$, of \cite{kg16}. Negative shifts of Fe\,I RV curves with respect to the $v_r^{\rm g}$ curve indicate an outflow of matter from the donor star. The Fe\,I 6469\,\AA\, line shows the highest outflow velocities, with an average value of $-3.8$\kms (dotted line). At $\varphi\approx 0$ - $0.2$, low inflow velocities are indicated for a fraction of Fe\,I lines. The data points are duplicated at orbital phases 1.0 - 2.0 for a more continuous view of the variability.
         }
\label{fig:RVfi}
\end{figure}

\begin{figure}
\centering
\begin{center}
\resizebox{\hsize}{!}{\includegraphics[angle=0]{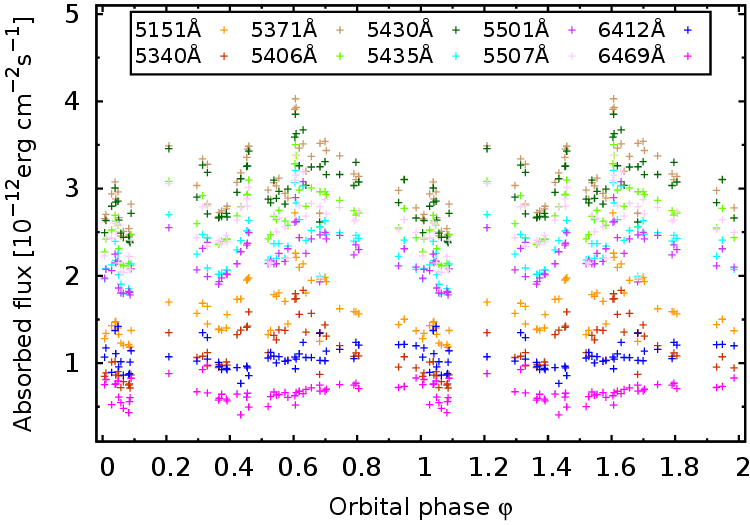}}
\end{center}
\caption{
Orbital variability of the Fe\,I absorption lines fluxes. The strongest absorption near the superior conjunction of the RG suggests the highest Fe\,I column densities between the RG and the apex of the neutral wind cone-shaped region (Sect.~\ref{s:int} and \ref{ss:orbvar}). The data points are duplicated at orbital phases 1.0 - 2.0 for more continuous view of the variability.
         }
\label{fig:FLfi}
\end{figure}

The resulting variability in RV values, $v_r$, is plotted in Fig.~\ref{fig:RVfi} together with the RV curve of the RG according to the solution of \cite{kg16}. Shifts up to $\approx -5$ \kms\ in the RVs of individual Fe\,I absorption lines relative to the RG curve are measured. This is consistent with a slow outflow of the absorbing material. However, around orbital phases $\varphi \approx 0$ - $0.2$, the RV values especially of the Fe\,I 5340\,\AA\ line suggest a slow inflow.

\begin{figure}
\centering
\begin{center}
\resizebox{8cm}{!}{\includegraphics[angle=0]{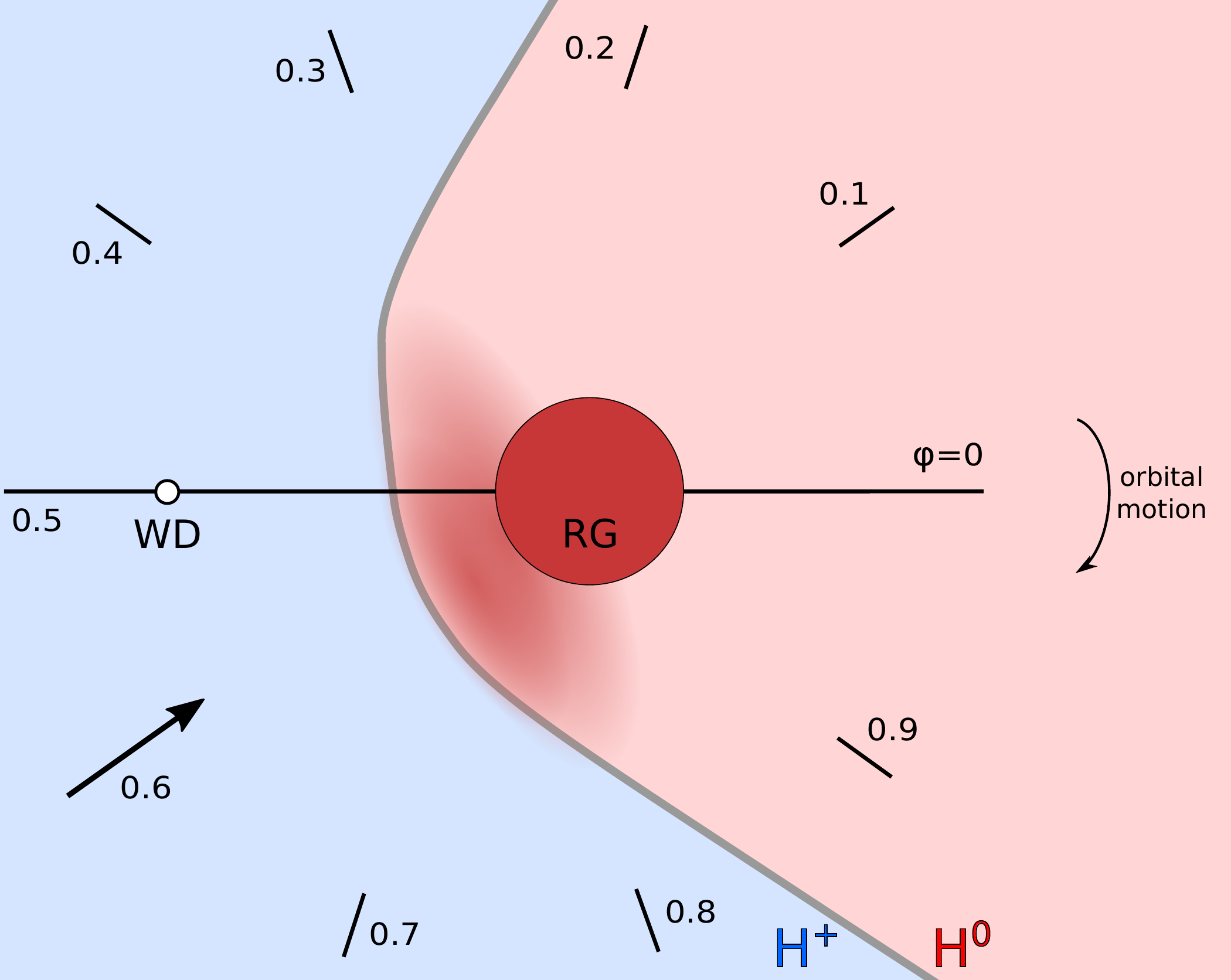}}
\end{center}
\caption{
Sketch of the ionization structure of \object{EG And} at the orbital plane \citep[adapted from][]{sh+21}. The density enhancement in the neutral wind zone is indicated in the direction towards the white dwarf, with the maximum column density at orbital phase $\approx 0.6$, causing the highest absorption in the Fe\,I lines (Sec.~\ref{ss:orbvar}).
}         
\label{fig:scheme}
\end{figure}

The orbital variability of the fluxes (Fig.~\ref{fig:FLfi}) shows the strongest absorption around the orbital phase $\approx 0.6,$ and the possible weakest absorption can be indicated at $\varphi\approx 0.1$, but there is a lack of the data around this orbital phase. When the conical shape of the neutral wind area around the RG is taken into account \citep[][ also Sect.~\ref{s:int}]{vo91}, this result points to the highest densities of the wind between the RG and the apex of the neutral area cone (Fig.~\ref{fig:scheme}). This agrees with the orbital variability of the absorption and the core-emission component of the H$\alpha$ line, which suggests that high-density matter lies in the area between the binary stellar components \citep{sh+21}. The complete list of measured RV and flux values is given in Tables~\ref{tabAp:RVs} and \ref{tabAp:fluxes} in the appendix.

\subsection{Model atmosphere grid}
\label{ss:mod}

The spread of RV values of the Fe\,I absorption lines within $\approx -5/+3$\kms around the RV curve of the RG, $v_r^{\rm g}$ (Fig.~\ref{fig:RVfi}) suggests that these lines originate in the vicinity of the stellar surface. To match the velocities with a depth in the RG atmosphere through modelling the profiles of Fe\,I lines (Sect.~\ref{ss:lprof}), we constructed a semi-empirical model atmosphere. This model is based on a simplified extension of the MARCS model atmosphere \citep{gu+08} up to a distance of 150\,$R_{\rm g}$ from the stellar centre. We defined the distribution of three physical parameters in the atmosphere as a logarithmically spaced grid: the neutral hydrogen density $N_{\rm H}$\,[cm$^{-3}$], temperature $T$\,[K], and electron pressure $P_{\rm e}$\,[Ba] over the required range of radial distance $r$\,[$R_{\rm g}$].

The MARCS model atmosphere extends up to a distance of 1.1\,$R_{\rm g}$ from the stellar centre. From the available database,\footnote{\url{https://marcs.astro.uu.se}} we selected the model with parameters closest to those of the RG in \object{EG And} (Sect.~\ref{s:int}), a moderately CN-cycled model with $^{12}$C/$^{13}$C$=20$, with a spherical geometry, effective temperature $T_{\rm eff}=3700$\,K, mass $M=1.0$\mo, $\log g = 0.5$, metallicity [Fe/H]\,$=0,$ and microturbulence parameter of 2\kms, which is a typical value for RGs in S-type SySts \citep{ga+16,ga+17}. The selected model atmosphere corresponds to a star with a radius $R = 93 $\ro\, and a luminosity $L = 1478 $\lo.

Beyond the radial distances covered by the MARCS atmosphere, we set the extrapolation up to $r=150$\,$R_{\rm g}$, where the wind density is sufficiently low to have a negligible impact on the Fe\,I line absorption profile. At this outer edge of the atmosphere model, we estimated values of $N_{\rm H}$ and $T$ from the hydrodynamical simulation of the M-giant $\gamma$\,Eri wind by \cite{wo+16}. We assessed the corresponding value of $P_{\rm e}$ for a representative value of the ionization fraction $\approx 10^{-6}$ for dense interstellar medium clouds \citep{wi+98,ca02,br+21}.

\begin{figure}
\centering
\begin{center}
\resizebox{\hsize}{!}{\includegraphics[angle=0]{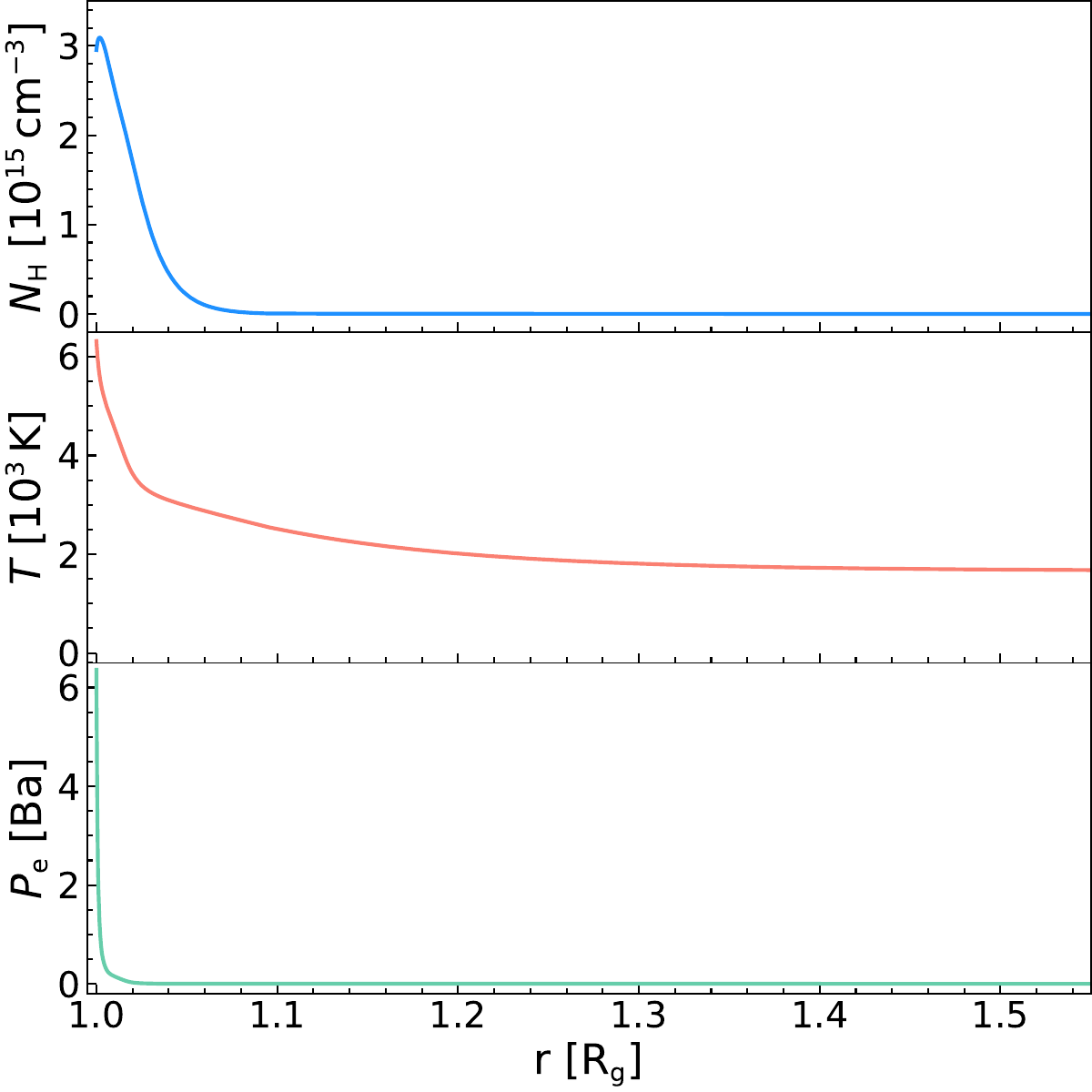}}
\end{center}
\caption{
Distribution of the neutral hydrogen density $N_{\rm H}$, temperature $T,$ and electron pressure $P_{\rm e}$ in the semi-empirical model atmosphere defined for the range $1$ - $150$\,$R_{\rm g}$ from the RG centre (Table~\ref{tab:atm} and Sect.~\ref{ss:mod}).
         }
\label{fig:atm}
\end{figure}

We defined the values of the physical parameters $N_{\rm H}$, $T$ and $P_{\rm e}$ between a radial distance 1.1 and 150\,$R_{\rm g}$ by interpolating the corresponding functions (Table~\ref{tab:atm}). The selection of the $N_{\rm H}(r)$ interpolation function has a crucial effect on the Fe\,I absorption line profile. We used the form corresponding to the model of measured H$^0$ column densities of \object{EG And} by \cite{sh+16},
\begin{equation}
   N_{\rm H}(r) = \displaystyle\frac{n_1}{2\lambda_1 R_{\rm g}}\frac{1 + \xi r^{1-K}}{r^2},
\label{eq:NHr}
\end{equation}
where $n_1$, $\xi$ \footnote{this parameter is given as $\xi=n_K\lambda_1/(n_1\lambda_K)$, where $n_K$ is a model parameter, and $\lambda_K$ is the $K$th eigenvalue of the Abel operator} and $K$ are the model parameters, and $\lambda_1 = \pi/2$ is the Abel operator eigenvalue \citep{kn+93}.
Since the column density model is most reliable at distances of $r$ of several $R_{\rm g}$, we applied the condition on the interpolation function (\ref{eq:NHr}) that $N_{\rm H}(r=3\,R_{\rm g})=1.6\times 10^{10}$\,cm$^{-3}$, that is, it equals the value of model J ($i=80^\circ$) from \cite{sh+16}. 
This approach led to smooth profiles of the atmosphere parameters over the required range of radial distances (Fig.~\ref{fig:atm}).

Finally, we took the asymmetric conical shape of the neutral wind zone into account. For orbital phases when the line of sight crosses the boundary between neutral and ionized wind, we estimated its distance from the RG surface from Fig.~6 in \cite{sh+21}. We assumed that only the neutral wind contributes to the absorption in Fe\,I lines. Therefore, we limited the radial size of the model atmosphere to the H$^0/$H$^+$ area border at these orbital phases. At the rest of the orbital phases, the radial length of the neutral area was assumed to be 150\,$R_{\rm g}$.

\begin{table*}[t!]
   \caption{
Model atmosphere. More details can be found in Sect.~\ref{ss:mod}. 
           }
\label{tab:atm}
\centering
\begin{tabular}{lcll}
\hline
\hline
\noalign{\smallskip}
  $r\,[R_{\rm g}]$  &   $1$ - $1.1$   & \hspace{3cm}$1.1$ - $150$ & \hspace{3cm}$150$ \\
\noalign{\smallskip}
\hline
\noalign{\smallskip}
  $N_{\rm H}$ & MARCS &   interpolation by function from \cite{sh+16} $^{a}$  & $10^{4}$\cmt\, \citep{wo+16} \\
  $T$         & MARCS &   interpolation by exponential + linear function  & $30\,$K\, \citep{wo+16} \\
  $P_{e}$     & MARCS &   interpolation by exponential function   & $10^{-17}\,$dyne/cm$^{2}$ (dense ISM clouds) $^{b}$ \\
\noalign{\smallskip}
\hline
\end{tabular}
\begin{flushleft}
{\bf Notes.}\\ 
$^{a}$\, with the condition that for $r=3\,R_{\rm g}$, $N_{\rm H}$ equals the value of model J from \cite{sh+16} \\
$^{b}$\, derived for a typical value of the ionization fraction of $10^{-6}$ \citep{wi+98,ca02,br+21}
\end{flushleft}
\end{table*}

\subsection{Line profile of the Fe\,I absorption lines}
\label{ss:lprof}

To reproduce the spectral profiles of ten Fe\,I absorption lines from 5151 to 6469\,\r{A}  at all orbital phases with a step of 0.1, we considered several broadening mechanisms that we incorporated into a custom Python code. We used the mass absorption coefficient including natural, pressure, thermal, and microturbulence broadening in the form given by \cite{gr05}.
The values of the Ritz wavelengths, the inner quantum numbers J, the oscillator strengths, and the excitation potentials were acquired from the National Institute of Standards and Technology (NIST) database\footnote{\url{https://www.nist.gov/pml/atomic-spectra-database}} and the natural damping constants from the Vienna Atomic Line Database\footnote{\url{ http://vald.astro.uu.se}} (VALD). The values of the partition functions for Fe\,I, Fe\,II, and Fe\,III were interpolated through the atmosphere grid from the tables of \cite{ha+02} and \cite{gr05}. For the atmosphere layers with temperatures below 1000\,K, we assumed constant partition functions.
We calculated the values of the Hjerting function as the real part of the Fadeeva function with the {\it wofz} function within scipy.special library\footnote{\url{ https://docs.scipy.org/doc/scipy/reference/generated/scipy.special.wofz.html}}.
The pressure broadening was treated as caused by the collisions with neutral hydrogen, using the impact approximation with line-broadening cross sections computed as a function of the effective principal quantum numbers \citep{an+95,ba+98} with the tabulated values of the broadening cross-section $\sigma$ and velocity parameter $\alpha$ given by \cite{ba+00}.

Furthermore, we included rotational broadening using the Python {\it rotBroad} function that is part of the PyAstronomy.pyasl library \footnote{ \url{https://pyastronomy.readthedocs.io/en/latest/pyaslDoc/aslDoc/rotBroad.html}}. Since the projected rotational velocity $v_{\rm rot} \sin (i)$ can be dependent on tidal forces in the outer regions of the RG \citep{mi+14}, we allowed it to be a free parameter. After first fitting trials with a free linear limb-darkening coefficient $\varepsilon$, most of the fits converged to $\varepsilon=1$. As this is a reasonable value \citep[tables of][]{ne+13}, we kept $\varepsilon=1$ in all line-profile fits.

As the typical value of the macroturbulence velocity in RGs is $\approx 3$\kms \citep{fe+03}, it adds to the broadening of the absorption-line profile. Often, the radial-tangential (RT) anisotropic macroturbulence is the preferred broadening model in a spectroscopic analysis \citep[e.g.][]{ca+08,si+17}.
On the other hand, \cite{ta+17} showed that the RT macroturbulence model is not adequate at least for solar-type stars because it overestimates turbulent velocity dispersion. They obtained more preferable results for the Gaussian anisotropic macroturbulence model. The resolution of our spectra and relatively low macroturbulent velocity does not allow us to distinguish between different macroturbulence models. Generally, there is agreement that neglecting macroturbulence as a source of line broadening leads to overestimated values of $v_{\rm rot} \sin (i)$, and, on the other hand, including a simple isotropic Gaussian macroturbulence model provides severely underestimated values of $v_{\rm rot} \sin (i)$ \citep{ae+09,si+14}. Therefore, we decided to include the isotropic Gaussian model with two values of macroturbulence velocity, 0 and 3\kms, to obtain lower and upper limits of $v_{\rm rot} \sin (i)$ values.

\begin{figure*}
\centering
\begin{center}
\resizebox{\hsize}{!}{\includegraphics[angle=0]{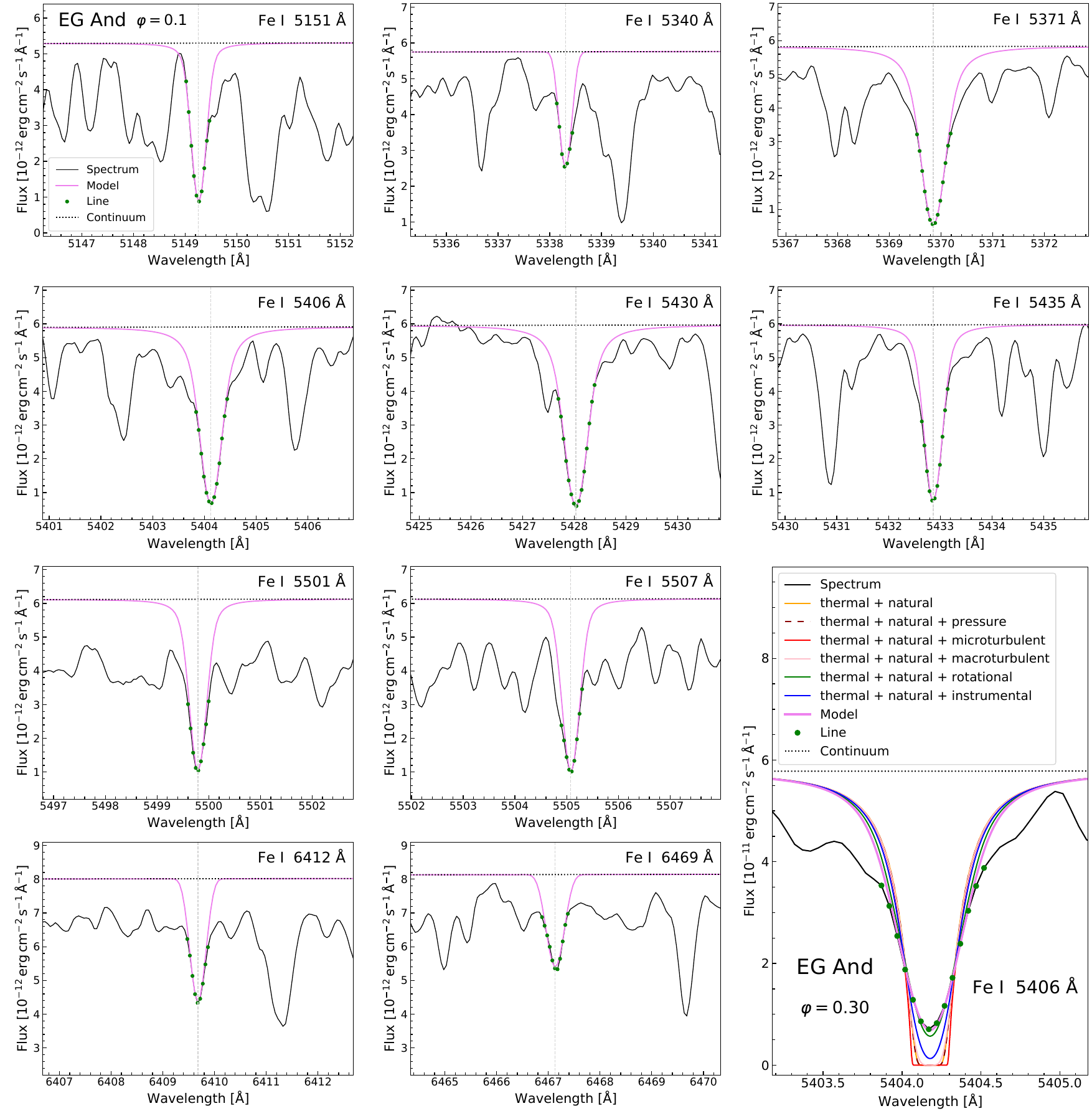}}
\end{center}
\caption{
Example line-profile fits at orbital phase $\varphi=0.1$ with a   macroturbulent velocity of 3\kms. The bottom right panel shows a comparison of the strength of different broadening mechanisms in the line-profile model for the Fe\,I 5406\,\AA\, line at $\varphi=0.3$.
         }
\label{fig:fits01}
\end{figure*}

Finally, we included the instrumental broadening using a Gaussian kernel. The width of the Gaussian profile used in the convolution is given by the resolution $R$, which depends on the wavelength and was estimated using ThAr lines.
For the wavelength range of the selected lines $5151-6469$\,\r{A}, the spectral resolution of our spectra ranges from $\approx 39100$ to $24000$. We used the {\it broadGaussFast} function from the PyAstronomy.pyasl library\footnote{ \url{https://pyastronomy.readthedocs.io/en/latest/pyaslDoc/aslDoc/broad.html}} to include macroturbulent and instrumental broadening. The line profiles depicted at the right bottom panel of Fig.~\ref{fig:fits01} compare the strength of individual broadening mechanisms.

We performed line profile modelling of Fe\,I lines at nine different orbital phases. Example fits at orbital phase $\varphi=0.1$ are depicted in Fig.~\ref{fig:fits01}. 
We evaluated the goodness of fit using the reduced $\chi$-square, $\chi_{\rm red}^2$. 
Its value is often $> 2$ due to the low value of the degrees of freedom and the uncertain value of the standard observational error, which can vary from one observation to the next. We adopted a rather strict value of a $2\%$ standard deviation of the flux values for all observations to avoid overestimating the errors for the best-quality spectra. 
The errors due to the simplified model of macroturbulence are relatively small (Fig.~\ref{fig:radvel} and \ref{fig:rotvel}) and have practically no effect on the resulting maximum depth of the origin of the spectral line in the atmosphere. The values of the $v_{\rm rot} \sin (i)$ parameter could be affected by unresolved blending of an absorption line.

An important source of systematic error can be introduced by the simplifications in our model, namely the symmetry of the wind distribution, which except for the shape of the neutral area, does not reflect the asymmetry of the  egress/ingress and orbital-plane/pole-region in the distribution of the physical quantities (Fig.~4 of \cite{du+99} and Fig.~8 of \cite{sh+21}). Moreover, the particular shape of the neutral area itself represents a further source of systematic error. We estimated the distance from the RG centre to the ionization boundary by adapting the shape of the neutral area computed for the symbiotic binary \object{SY Mus} \citep{sh+17,sh+21}. This system comprises a white dwarf that is more luminous than the hot companion in \object{EG And by two orders of magnitude} \citep{mu+91}. On the other hand, the mass loss from its RG is probably higher \citep{mu+91,sk05}, leading to a denser wind zone. These characteristics affect the shape of the ionization boundary, and the actual shape for \object{EG And} can therefore deviate from the one in \object{SY Mus}. However, the similar measured egress values of the H$^0$ column densities and the practically identical asymptote to the egress ionization boundary in the orbital plane, which is located at $\varphi\approx 0.17$ \citep[Fig. 2b of][]{sk23}, strongly suggest that the ionization boundaries in these systems are similar.
The lack of measured H$^0$ column densities at ingress orbital phases for \object{EG And} precludes us from modelling the full shape of the ionization boundary. To estimate the sensitivity of the resulting physical parameters on the location of the ionization boundary, we performed fits for a shifted radial distance of the ionization boundary by $-0.5R_{\rm g}$ and $+0.5R_{\rm g}$ for a subset of modelled spectra with all Fe\,I absorption lines and orbital phases with the finite radial size of the neutral area represented. For the ionization boundary closer to the RG by $-0.5R_{\rm g}$, we obtained the same values of column densities or lower values by up to 6.0\%, and for the boundary that is more distant by $+0.5R_{\rm g}$, the values were the same or higher by up to 1.1\%. In both cases, higher values of the errors of $n_{\rm H}$ correspond to orbital phases $\approx$ 0.4 - 0.7, where the position of the ionization boundary is closer to the RG. The corresponding values of the projected rotational velocity $v_{\rm rot} \sin (i)$ remained unchanged for all fits, as did the values of the minimum distance from the RG centre $r$ (Sect.~\ref{sss:hephot}). This confirms the dominant role of the densest parts of the RG atmosphere in the formation of Fe\,I absorption line profiles. Given the rather low magnitude of the errors of $n_{\rm H}$ due to the uniform shifts and the most probably similar shape of the ionization structure for both systems, which is supported by similar profiles of the measured H$^0$ column densities \citep{sh+16}, an uncertain precise location of the apex of the neutral zone will probably not seriously affect the ratios of $n_{\rm H}$ values at individual orbital phases yielded by the line-profile modelling.

Another source of systematic error comes from the uncertain level of the continuum, which is mainly due to the spread in the photometric data. In our dataset, the typical deviation of the continuum values from the average relative to the flux ranges from 3\% to 9\% at the positions of individual Fe\,I absorption lines. This leads to errors in the $n_{\rm H}$ values with a magnitude of typically $\approx 10$ - $20$\%, $v_{\rm rot} \sin (i)$ of $\approx 1$ - $7$\% and a minimum distance $r$ of $\approx 0.1$ - $0.5$\%. Therefore, the uncertainty in the level of the continuum represents a more significant source of error than the uncertainty in the position of the ionization boundary. Still, these systematic errors are of lower magnitude than the values of the standard deviations of the resulting values from the set of modelled spectra.

\subsection{Distribution of the physical parameters within the atmosphere}
\label{ss:strat}

\subsubsection{The height above the photosphere}
\label{sss:hephot}

\begin{figure}
\centering
\begin{center}
\resizebox{\hsize}{!}{\includegraphics[angle=0]{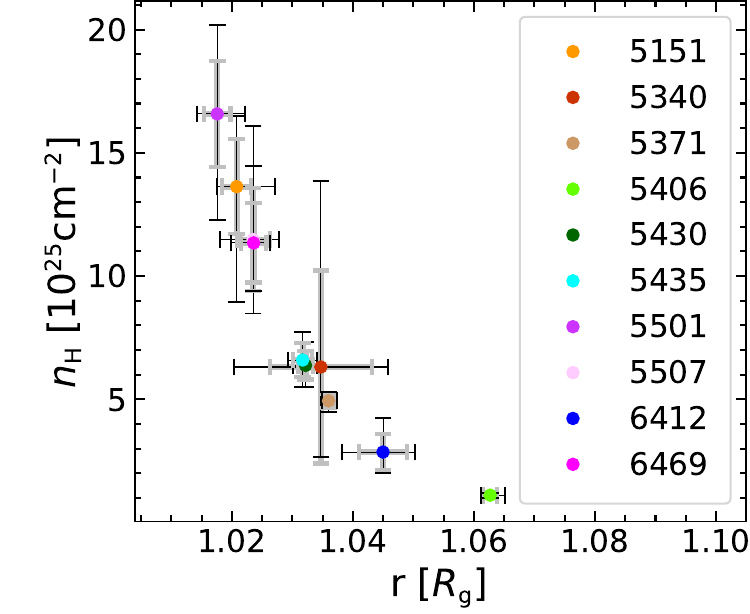}}
\end{center}
\caption{
Column density as a function of the average minimal distance $r$ from the RG centre over our set of orbital phases and models with and without macroturbulence for individual Fe\,I absorption lines (dots), based on the values in Table \ref{tab:averpar}. 
 The grey error bars show the standard deviation, and the black error bars represent 
 the total range of determined $r$ from the models. 
         }
\label{fig:rLines}
\end{figure}

\begin{figure*}
\centering
\begin{center}
\resizebox{13cm}{!}{\includegraphics[angle=0]{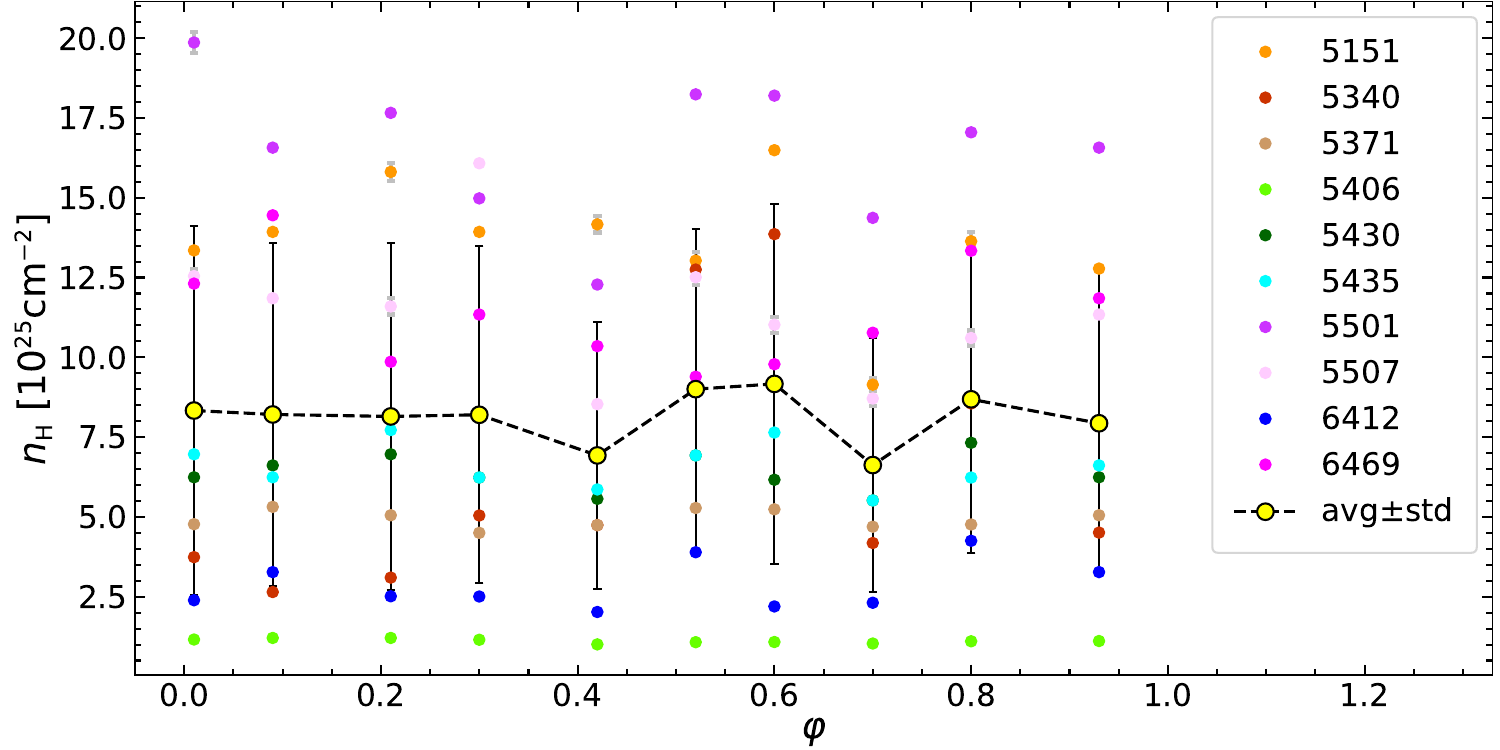}}
\end{center}
\caption{
 $n_{\rm H}$ values as a function of orbital phase for individual Fe\,I absorption lines and their averages. For most values, the differences between  the models with and without macroturbulence are smaller than the size of the data points. When they are larger, they are shown as grey error bars.
         }
\label{fig:nHfi}
\end{figure*}

Our models provided us with the total columns of the wind material that form the spectral profiles of individual Fe\,I lines in our set. From now on, the values of $r$ are understood as the distances of the lowest layers of the atmosphere model, corresponding to the resulting neutral columns from the line-profile fits. In other words, a particular value $r$ represents the maximum depth within the model atmosphere where the integration of the line-profile stops, and it corresponds to the deepest layer of the origin of the spectral line.
 The maximum depths of the Fe\,I line profile fits correspond to 
 a relatively small height, $\approx0.02$ to $\approx0.06$\,$R_{\rm g}$, above the RG photosphere. 
 Figure~\ref{fig:rLines} shows this result with the corresponding column 
 densities. The resulting physical parameters averaged over the orbital 
 phases are presented in Table~\ref{tab:averpar}. 
There is no sign 
of significant variations in the column density with orbital phase, but a 
slightly higher average value is measured at $\varphi=0.5-0.6$ (Fig.~\ref{fig:nHfi}).

%
%
\begin{table*}[t!]
   \caption{
Physical parameters $T$, $N_{\rm H}$, $n_{\rm H}$, $v_r$ , and $v_{\rm rot}\sin(i)$ averaged over $\varphi$ and their standard deviations, with the distance $r$ from the RG centre, corresponding to the lowest layer in the atmosphere of the total absorbing column for individual Fe\,I lines. The RV values averaged over all 53 observations (Sect.~\ref{s:obs}), $v^{\rm tot}_r$, are shown for comparison.
           }
\label{tab:averpar}
\centering
\begin{tabular}{cclrrrrr}
\hline
\hline
\noalign{\smallskip}
Fe\,I line                                  & 
$r$                                         & 
\multicolumn{1}{c}{$T$}                     &
\multicolumn{1}{c}{$N_{\rm H}$}             & 
\multicolumn{1}{c}{$n_{\rm H}$}             &
\multicolumn{1}{c}{$v_r$}                   & 
\multicolumn{1}{c}{$v^{\rm tot}_r$}         & 
\multicolumn{1}{c}{$v_{\rm rot} \sin (i)$}  \\ 
\hspace{-0.6mm} [\AA]                       &
\,[$R_{\rm g}$]                             &
\multicolumn{1}{c}{[K]}                     &
\multicolumn{1}{c}{[10$^{14}$cm$^{-3}$]}    &
\multicolumn{1}{c}{[10$^{25}$cm$^{-2}$]}    &
\multicolumn{3}{c}{[\kms]}                  \\
\noalign{\smallskip}
\hline
\noalign{\smallskip}
     5151 & 1.021 $\pm$ 0.002 & 3599 $\pm$ 123 & 16.4 $\pm$ 2.0 & 13.6 $\pm$ 1.9 & -0.37 $\pm$ 0.36 & -0.29 $\pm$ 0.82 &  9.68 $\pm$ 1.05 \\
     5340 & 1.035 $\pm$ 0.008 & 3216 $\pm$ 198 &  7.7 $\pm$ 4.8 &  6.3 $\pm$ 3.9 &  0.59 $\pm$ 0.62 &  0.54 $\pm$ 0.83 & 12.17 $\pm$ 2.08 \\
     5371 & 1.036 $\pm$ 0.001 & 3149 $\pm$  13 &  6.1 $\pm$ 0.4 &  4.9 $\pm$ 0.3 & -0.42 $\pm$ 0.62 & -0.38 $\pm$ 0.84 & 12.83 $\pm$ 0.99 \\
     5406 & 1.063 $\pm$ 0.001 & 2849 $\pm$  11 &  0.8 $\pm$ 0.1 &  1.1 $\pm$ 0.1 & -0.44 $\pm$ 0.55 & -0.45 $\pm$ 0.90 & 11.15 $\pm$ 0.76 \\
     5430 & 1.032 $\pm$ 0.001 & 3212 $\pm$  24 &  8.0 $\pm$ 0.7 &  6.4 $\pm$ 0.6 & -0.95 $\pm$ 0.55 & -0.82 $\pm$ 0.91 & 12.24 $\pm$ 0.47 \\
     5435 & 1.032 $\pm$ 0.002 & 3222 $\pm$  30 &  8.3 $\pm$ 0.9 &  6.6 $\pm$ 0.7 & -0.65 $\pm$ 0.44 & -0.58 $\pm$ 0.92 & 10.49 $\pm$ 1.13 \\
     5501 & 1.018 $\pm$ 0.002 & 3823 $\pm$ 169 & 19.1 $\pm$ 1.8 & 16.6 $\pm$ 2.1 & -0.22 $\pm$ 0.64 & -0.10 $\pm$ 0.92 & 11.10 $\pm$ 1.00 \\
     5507 & 1.024 $\pm$ 0.003 & 3468 $\pm$ 125 & 14.1 $\pm$ 2.2 & 11.5 $\pm$ 2.1 & -2.08 $\pm$ 0.57 & -2.10 $\pm$ 0.90 & 10.78 $\pm$ 1.37 \\
     6412 & 1.045 $\pm$ 0.004 & 3034 $\pm$  47 &  3.3 $\pm$ 1.0 &  2.9 $\pm$ 0.7 & -0.74 $\pm$ 0.56 & -0.82 $\pm$ 0.85 &  9.62 $\pm$ 2.12 \\
     6469 & 1.024 $\pm$ 0.002 & 3456 $\pm$  92 & 14.0 $\pm$ 1.7 & 11.3 $\pm$ 1.6 & -3.65 $\pm$ 0.87 & -3.79 $\pm$ 0.87 &  8.48 $\pm$ 3.09 \\
\noalign{\smallskip}
\hline
\end{tabular}
\end{table*}

\subsubsection{Radial velocities}
\label{sss:radvel}

\begin{figure}
\centering
\begin{center}
\resizebox{\hsize}{!}{\includegraphics[angle=0]{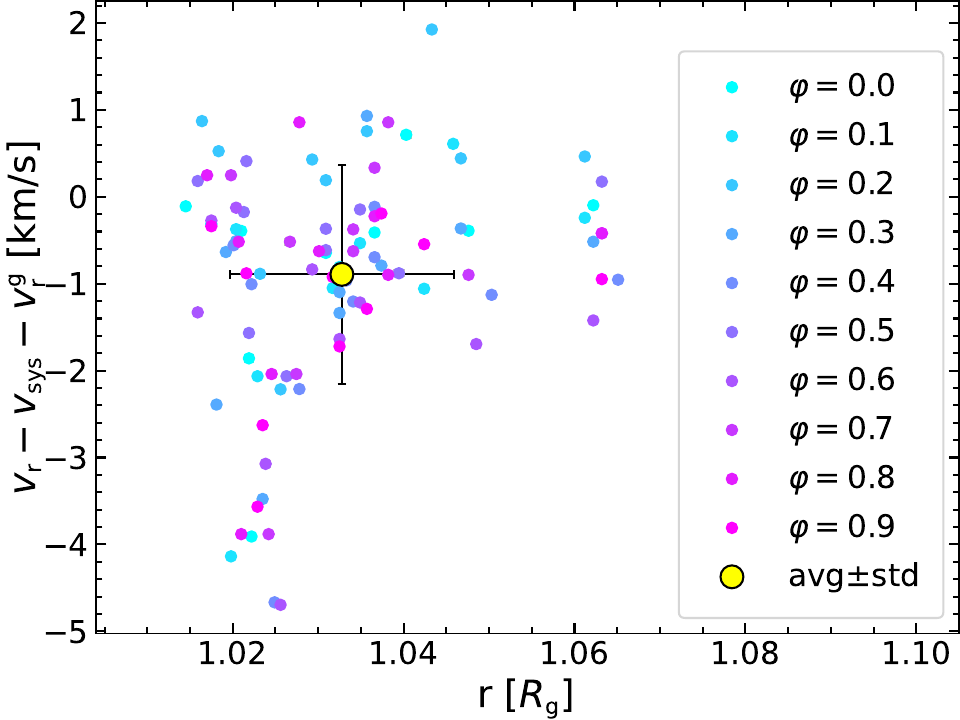}}
\end{center}
\caption{
Distribution of RVs, corrected for the systematic velocity and orbital motion of the RG, along the radial direction at different orbital phases.
The differences between  models with and without macroturbulence are smaller than the size of data points.
The large yellow circle corresponds to the average value of all RV values, and the black error bars show the standard deviation.
         }
\label{fig:radvel}
\end{figure}

\begin{figure}
\centering
\begin{center}
\resizebox{\hsize}{!}{\includegraphics[angle=0]{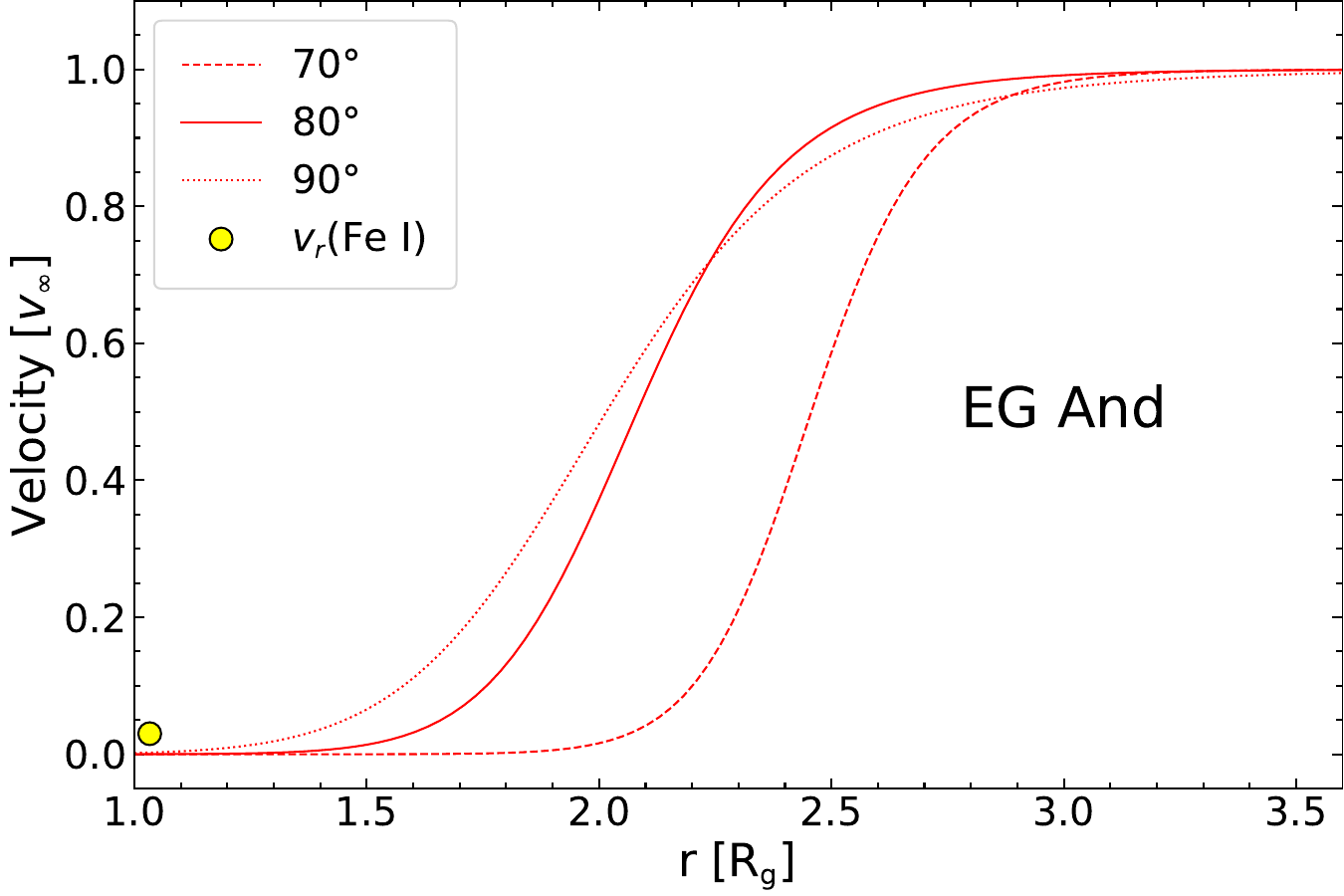}}
\resizebox{\hsize}{!}{\includegraphics[angle=0]{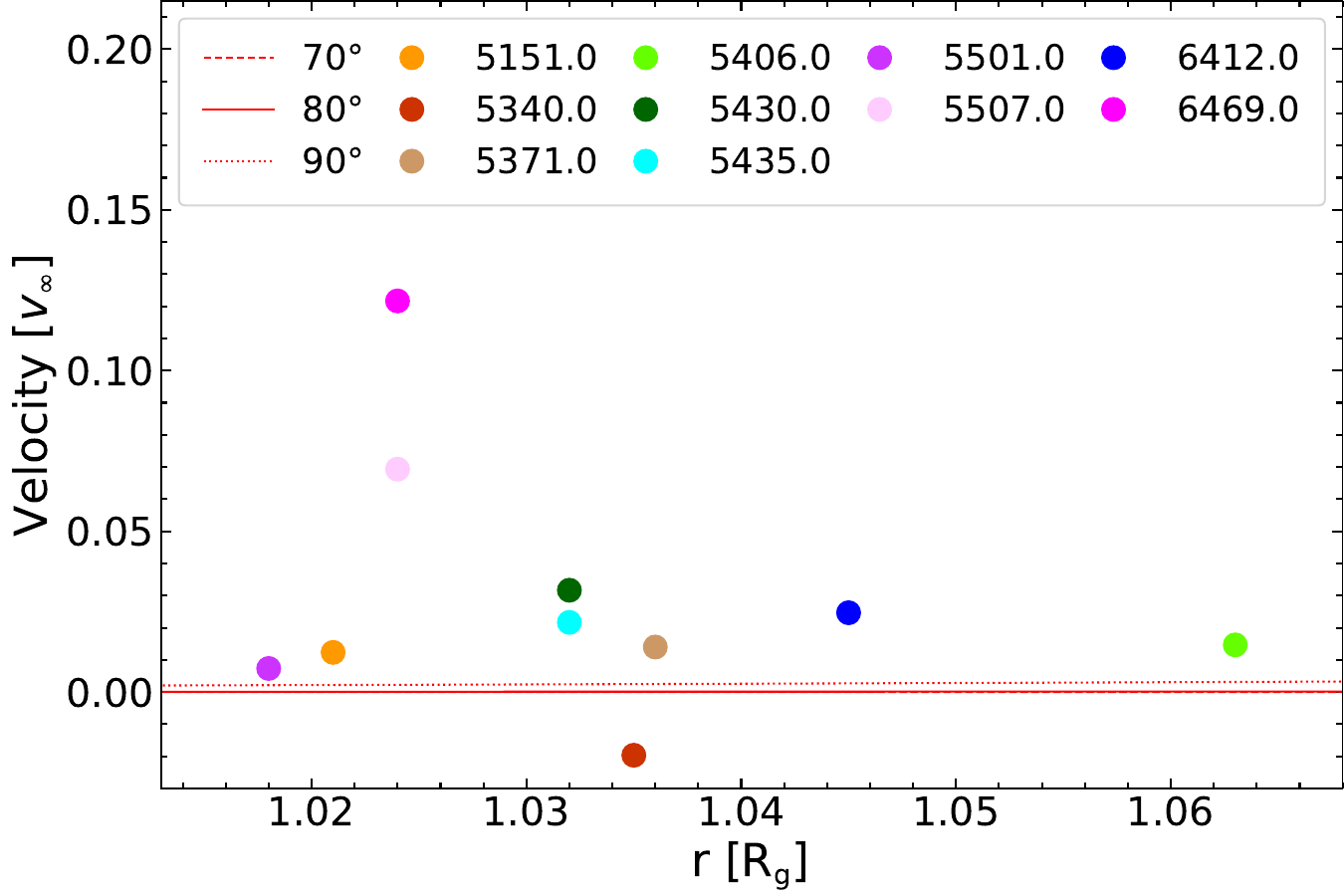}}
\end{center}
\caption{
Comparison of the wind velocity profiles of RG in \object{EG And} \citep{sh+16} with the RV values of the Fe\,I absorption lines, $v_r(\rm Fe\,I)$, that were paired with the distance $r$ by the line-profile modelling (Sect.~\ref{ss:lprof}). In the top panel, the $v_r(\rm Fe\,I)$ data point represents the average value over all modelled spectra (Sect.~\ref{sss:radvel}). The bottom panel is a zoomed view with $v_r(\rm Fe\,I)$ values for the individual Fe\,I lines averaged over the orbital phases.
         }
\label{fig:vrlaw}
\end{figure}

\begin{figure}
\centering
\begin{center}
\resizebox{\hsize}{!}{\includegraphics[angle=0]{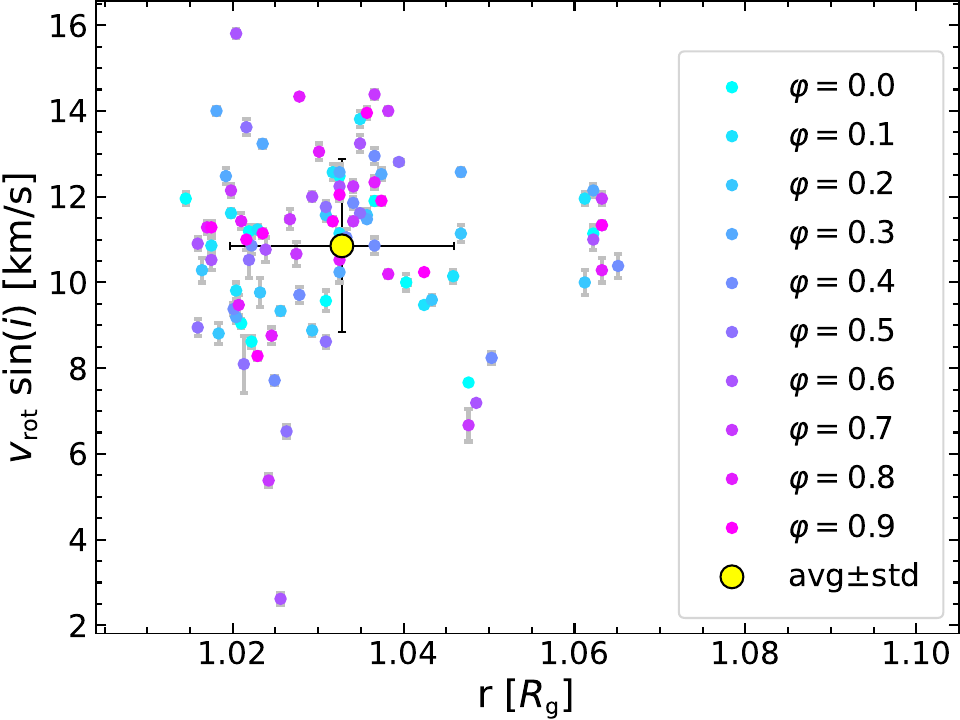}}
\end{center}
\caption{
As in Fig.~\ref{fig:radvel}, but for projected rotational velocity.
The grey error bars correspond to the difference between the models with and without macroturbulent broadening. Where no error bar is visible, the difference is smaller than the size of a data point.
         }
\label{fig:rotvel}
\end{figure}

The total average and standard deviation over ten modelled spectra and ten Fe\,I absorption lines corresponds to RV $-0.89\pm 1.26$\kms at a radial distance $1.03\pm 0.01$\,$R_{\rm g}$ (Fig.~\ref{fig:radvel}). Assuming a terminal velocity of 30\kms, we compared our RV values with velocity profiles obtained for \object{EG And} from modelling the measured column densities by \cite{sh+16}. As shown in Fig.~\ref{fig:vrlaw}, our results support very slow wind velocities close to the RG surface before the acceleration of the wind starts.

\subsubsection{Rotational velocities}
\label{sss:rotvel}

The orbit-averaged values of the projected rotational velocities of all modelled lines fall within 9.6 - 12.8\kms\, with standard deviations of 4 - 22\% (Table~\ref{tab:averpar}), except for the Fe\,I 6469\,\AA\, line with $v_{\rm rot} \sin (i) = 8.5$ and a significantly higher standard deviation of 36\%.
While it is reasonable not to expect the same rotational velocity in any depth in the RG atmosphere, the measured differences can in part be caused by errors due to the blending of the lines. Moreover, the reliability of the $v_{\rm rot} \sin (i)$ determination is affected by the comparable strength of the instrumental broadening.
The average and standard deviation over the whole sample of ten line-profile models per ten fitted lines corresponds to $v_{\rm rot} \sin (i) = 10.9 \pm 2.0$\kms, which is in a typical range of $\approx 5$ - $11$\kms determined for RGs in S-type SySts \citep{za+08,ga+16,ga+17}. There are also much faster rotators in this group of stars with $v_{\rm rot} \sin (i)$ up to $\approx 50$\kms\,\citep{za+07}.

Assuming an orbital inclination of $i=80^\circ\pm 10^\circ$, we obtained $v_{\rm rot} = 11.1_{-2.2}^{+2.6}$\kms. Then, for RG radius $R_{\rm g}=75\pm 10$\,$R_\odot$, the proportion of orbital to rotational period is $P_{\rm orb}/P_{\rm rot}=1.4_{-0.4}^{+0.6}$. Therefore, it is possible that the rotation of the RG is bounded to its orbital motion.

\section{Discussion}
\label{s:dis}

For our sample of ten Fe\,I absorption lines, we determined 
the absorbed flux and RV from their Gaussian fits 
(Sect.~\ref{ss:orbvar}). Both quantities show the relative 
displacements for individual lines along the orbit 
(Figs.~\ref{fig:RVfi} and \ref{fig:FLfi}). 
The largest average shift in RVs by $-3.8$\kms\ with respect 
to the $v_r^{\rm g}(\varphi)$ curve is shown by the 
Fe\,I\,6469\,\AA\ line (Fig.~\ref{fig:RVfi}, dotted line). 
Around $\varphi = 0.1$, the RVs of many lines indicate 
a slow flow of absorbing material towards the RG, especially the Fe\,I 5340\,\AA\ line with an average RV shift of $+0.5$\kms.
 
The line-profile models accounting for several broadening 
mechanisms at the selected ten orbital phases (Sect.~\ref{ss:lprof}) 
enabled us to match their RV values with the deepest layer of 
the atmosphere, where the absorption line is predominantly 
created, characterized by $r$, $T$, $N_{\rm H}$ and $P_{\rm e}$ 
values. For the resulting depths in the range of
$1.02$ - $1.06$ \,$R_{\rm g}$, all averaged RV values are low. 
Specifically, the outflow values lie within the interval from $-0.2$ to 
$-3.7$\kms\, (Table~\ref{tab:averpar}). This represents 
$0.7$ - $12.3\,\%$ of the estimated terminal wind velocity of 
$30$\kms. While the typical RV at $r \approx 1.03$\,$R_{\rm g}$ 
is $\approx 1$\kms\, ($\approx 3\%$  of $v_\infty$), there is 
a considerable dispersion in individual RV values (Fig.~\ref{fig:vrlaw}, bottom).
The highest range of RV values is measured at the shortest 
distances of $\approx 1.02$ - $1.03$\,$ R_{\rm g}$. This variability 
can be a result of the highly complex flows of matter in the close 
surroundings of cool evolved stars \citep{kr+18}.

In the light of our results, the orbital phase $\approx 0.6$ 
seems to be exceptional in several ways. First, most of 
the Fe\,I lines from our set reach the maximum absorbed flux 
at this orbital phase (Fig.~\ref{fig:FLfi}), pointing to a higher column density in the 
neutral zone between the apex of its cone and the RG, that is, in the direction towards the white dwarf companion (Fig.~\ref{fig:scheme}). In the same way, 
we could interpret the local maxima in the resulting column densities 
of the line-profile models (Fig.~\ref{fig:nHfi}). 
Simultaneously, a higher dispersion of the RV values and 
the overall highest outflow velocities were measured around 
this orbital phase, suggesting enhanced outflow of the wind. 
The same feature was observed for the core-emission and 
absorption components of the H$\alpha$ line at orbital 
phases $0.6-0.7$ \citep{sh+21}.
Higher densities and, at the same time, higher velocities 
of the neutral matter may 
represent a challenge for hydrodynamical simulations of 
outflows from evolved cool stars in binary systems. 

In our previous work, we investigated the geometrical 
distribution of the RG wind in \object{EG And}. By 
modelling H$^0$ column densities, we found that the 
wind from the RG is focused towards the orbital 
plane \citep{sh+16}. On the other hand, the RV orbital 
variability of the [OIII]\,5007\,\AA\, line, which coincides with the $v_r^{\rm g}(\varphi)$ curve in both phase and amplitude, indicates a dilution of the 
wind around the poles of the RG \citep{sh+21}. However, the underlying mechanism that focuses wind in 
this system remains unclear. \cite{sk+15} 
applied the wind-compression disk model proposed by \cite{bc93} to RGs in S-type symbiotic systems with rotational velocities of 6-10\kms
and found that the wind 
focusing occurs at the equatorial plane with a factor 
of 5–10 relative to the spherically symmetric wind. 
The average 
value $v_{\rm rot} = 11.1$\kms (Sect.~\ref{sss:rotvel}) 
is therefore sufficiently high for rotation-induced compression of the wind from the giant in \object{EG And}.

The wind focusing can also potentially explain the higher 
densities of the neutral wind between the binary components, in contrast to the lower densities in the opposite direction, even though the neutral zone is more extended there.
However, the wind compression by the RG rotation cannot explain this asymmetry because this mechanism acts equally strongly in all outward directions in the plane perpendicular to the rotational axis.
Therefore, the gravitational effect of the white dwarf companion is the more natural explanation for this measured asymmetry.
In a recent 3D hydrodynamical simulation of the accretion process for representative parameters of S-type symbiotic systems by \cite{le+22}, the centre of the oblique region with highest densities around the RG is shifted towards the white dwarf, and the wind enhancement in the area of the orbital plane is also visible in their Fig.~2. For S-type system, recurrent nova \object{RS Oph}, the simulations of \cite{bo+16} showed a dense equatorial outflow in the system as a result of the interaction of a slow wind with a binary companion. Therefore, gravitational focusing likely shapes the circumstellar matter in S-type SySts, as well as in D-type systems \citep{vb+09}.

Often, the analysis of spectral lines in stellar atmospheres is focused on the determination of elemental abundances and basic stellar parameters by comparing synthetic and observational spectra \citep[e.g.][]{st+20,fu+21} In our work, we aimed to assess the physical conditions at different heights in the RG atmosphere in interacting binary star from Fe\,I absorption line profiles. In principle, this approach can also be used for isolated non-dusty RG stars, which can potentially have different wind velocity profiles.
The presence of the companion of a mass-loosing star affects the flow of matter in the wind region. Its gravitational pull can support the wind outflow from the RG, and we cannot exclude that in the case of single RGs, the low-velocity region is more extended and the velocities are lower. To form an idea about the proportions of gravitational force of the two stellar components in \object{EG And}, we compared the values of the gravitational force of the white dwarf and RG at several distances $r$ on the line joining the two stars.
When we assume the separation between the two components of $4.5R_{\rm g}$ from the interval given by \cite{vo91}, the magnitude of the white dwarf force at $r = 1.02 - 1.06\,R_{\rm g}$, where the Fe\,I absorption lines are predominantly created, is small but not negligible. It is about $2$\% of the value of the RG gravitational force. At $r = 1.5R_{\rm g}$, where the acceleration of the wind starts (Fig.~\ref{fig:vrlaw}, top), this value is $\approx 7$\%, and at $r = 2R_{\rm g}$ in the acceleration region, it is $\approx 17$\%. At the location at $\approx 3R_{\rm g}$, where the terminal velocity of the wind is reached, the gravitational forces from the two stars are already comparable. Close to the RG surface, where most of the absorption in Fe\,I lines occurs, the gravitational effect of the white dwarf is small, and we do not observe any tendency in the wind RVs as a function of orbital phase (Fig.~\ref{fig:radvel}), that is, at different distances of the near-surface regions from the white dwarf
companion. Therefore, the RVs near the surface of the RG in \object{EG And} are probably comparable to those in isolated giants with similar evolutionary and physical characteristics. In the future, modelling of the Fe\,I absorption line-profiles for single late-type giants can be used to probe this assumption.

\section{Conclusions}
\label{s:concl}

The RVs of the investigated Fe\,I absorption lines trace the orbital motion of the giant in the binary star \object{EG And}. They are displaced from the RV curve of the giant by 0.1 to 3.8\kms\ (i.e. up to 13\% of the terminal wind velocity), which indicates a slow outflow of mass from the RG (Fig.~\ref{fig:RVfi}).
Modelling of their profiles showed that they are formed at maximum depths from $\approx 0.02$ to $\approx 0.06$\,$R_{\rm g}$ above the photosphere. 
The typical value of the RV at these distances is around 1\kms , which is consistent with the previously determined wind velocity profile from measured values of H$^0$ column densities (Fig.~\ref{fig:vrlaw}). 
It is interesting to note that several Fe\,I lines, especially the 5340\,\AA\ line, showed a slow inflow of the absorbing matter towards the RG around orbital phase 0.1. Together with the dispersion of the RV values of several \kms, this may be a sign that the nature of the near-surface mass flows in the RG atmosphere is complex (Fig.~\ref{fig:RVfi} and \ref{fig:vrlaw}, bottom).

The orbital variations of the Fe\,I absorption line fluxes (Fig.~\ref{fig:FLfi}) indicate that higher-density matter resides in the region between the binary components than in other directions from the RG at the near-orbital plane area. This asymmetry can be the result of gravitational interaction of the white dwarf with the RG wind, as was indicated by numerical simulations of gravitationally focused winds in interacting binaries. 
The measured rotational velocity of the RG, $\approx 11.1$\kms, suggests an additional compression of the wind from the giant towards the orbital plane due to its rotation. Our results therefore support the contribution of both mechanisms to the observed RG wind enhancement and its asymmetry in the orbital plane of \object{EG And}.

The results of measuring the wind density asymmetry in the near-orbital plane region are consistent with our previous results on the wind focusing \citep{sh+16,sh+21}. Our direct observational finding shows a wind density enhancement between the binary components. This confirms the high efficiency of the wind mass transfer in SySts.


\begin{acknowledgements}
We wish to thank to Zoltán Garai, Andrii Maliuk, Matej Sekeráš and Peter Sivanič 
for obtaining 1-2 spectral/photometric observations each, used in this work. 
We acknowledge with thanks the variable star observations from 
the AAVSO International Database contributed by observers 
worldwide and used in this research.
This work was supported by the Slovak Research and Development
Agency under the contract No. APVV-20-0148 and by a grant of 
the Slovak Academy of Sciences, VEGA No. 2/0030/21. 
VK acknowledges the support from the Government Office of the Slovak Republic within NextGenerationEU programme under project No. 09I03-03-V01-00002.
Reproduced with permission from Astronomy \& Astrophysics, \copyright\,ESO.
\end{acknowledgements}


\bibliographystyle{aa}

\bibliography{reflist.bib}


\begin{appendix}

\onecolumn

\section{Radial velocities and fluxes of selected Fe\,I absorption lines}
\label{apx}


\begin{table*}[h]
\small
\setlength{\tabcolsep}{2pt}
   \caption{
Radial velocities [\kms] of ten Fe\,I absorption
lines from our 53 spectra (Sect.~\ref{ss:orbvar}). 
           }
\label{tabAp:RVs}
\centering
\begin{tabular}{cccrrrrrrrrrr}
\hline
\hline
\noalign{\smallskip}
 HJD      &     dateobs      &  phase  &  5151\,\AA  &    5340\,\AA &    5371\,\AA &    5406\,\AA &    5430\,\AA &    5435\,\AA &    5501\,\AA &    5507\,\AA &    6412\,\AA &   6469\,\AA \\
\noalign{\smallskip}
\hline
\noalign{\smallskip}
  2457743.439  &   2016/12/20.939  &  0.630  & $-100.433$ & $ -99.564$ & $-101.119$ & $-101.663$ & $-101.064$ & $-101.393$ & $-100.912$ & $-103.576$ & $-101.377$ & $-104.652$  \\
  2457914.508  &   2017/06/10.008  &  0.985  & $ -96.228$ & $ -95.340$ & $ -96.295$ & $ -96.376$ & $ -96.856$ & $ -96.584$ & $ -96.032$ & $ -97.854$ & $ -96.561$ & $ -99.639$  \\
  2457924.512  &   2017/06/20.012  &  0.006  & $ -95.142$ & $ -94.152$ & $ -94.849$ & $ -94.929$ & $ -95.397$ & $ -95.054$ & $ -94.684$ & $ -96.723$ & $ -95.791$ & $ -99.181$  \\
  2457926.523  &   2017/06/22.023  &  0.010  & $ -94.878$ & $ -93.954$ & $ -94.739$ & $ -94.764$ & $ -95.240$ & $ -95.064$ & $ -94.618$ & $ -96.451$ & $ -94.627$ & $ -98.536$  \\
  2457941.517  &   2017/07/07.017  &  0.041  & $ -94.165$ & $ -93.265$ & $ -94.056$ & $ -94.323$ & $ -94.618$ & $ -94.376$ & $ -93.762$ & $ -95.860$ & $ -94.471$ & $ -97.280$  \\
  2457964.549  &   2017/07/30.049  &  0.089  & $ -91.599$ & $ -90.550$ & $ -91.390$ & $ -91.409$ & $ -91.908$ & $ -91.822$ & $ -91.277$ & $ -93.156$ & $ -91.831$ & $ -95.114$  \\
  2458073.421  &   2017/11/15.921  &  0.314  & $ -88.743$ & $ -87.923$ & $ -88.748$ & $ -88.816$ & $ -89.213$ & $ -89.003$ & $ -88.640$ & $ -90.385$ & $ -88.948$ & $ -91.631$  \\
  2458080.387  &   2017/11/22.887  &  0.329  & $ -87.896$ & $ -87.388$ & $ -88.135$ & $ -87.856$ & $ -88.371$ & $ -88.272$ & $ -87.843$ & $ -89.656$ & $ -88.341$ & $ -91.072$  \\
  2458370.558  &   2018/09/09.058  &  0.930  & $ -98.878$ & $ -98.129$ & $ -99.020$ & $ -99.098$ & $ -99.458$ & $ -99.183$ & $ -98.432$ & $-100.505$ & $ -98.778$ & $-101.742$  \\
  2458900.210  &   2020/02/20.710  &  0.028  & $ -92.395$ & $ -91.631$ & $ -92.466$ & $ -92.443$ & $ -92.727$ & $ -92.389$ & $ -92.065$ & $ -94.140$ & $ -92.928$ & $ -96.204$  \\
  2458911.235  &   2020/03/02.735  &  0.050  & $ -91.755$ & $ -90.389$ & $ -91.513$ & $ -91.820$ & $ -92.031$ & $ -91.805$ & $ -91.350$ & $ -93.324$ & $ -91.467$ & $ -95.412$  \\
  2458914.242  &   2020/03/05.742  &  0.057  & $ -91.628$ & $ -90.462$ & $ -91.990$ & $ -92.019$ & $ -92.125$ & $ -91.962$ & $ -91.330$ & $ -93.651$ & $ -91.889$ & $ -95.305$  \\
  2458918.238  &   2020/03/09.738  &  0.065  & $ -90.959$ & $ -89.675$ & $ -91.254$ & $ -91.163$ & $ -91.350$ & $ -91.065$ & $ -90.779$ & $ -92.998$ & $ -91.417$ & $ -94.821$  \\
  2458926.248  &   2020/03/17.748  &  0.082  & $ -89.581$ & $ -89.041$ & $ -89.447$ & $ -89.626$ & $ -89.739$ & $ -89.666$ & $ -89.026$ & $ -91.350$ & $ -90.113$ & $ -94.362$  \\
  2458927.243  &   2020/03/18.743  &  0.084  & $ -89.551$ & $ -88.608$ & $ -89.415$ & $ -89.406$ & $ -89.707$ & $ -89.519$ & $ -88.990$ & $ -91.125$ & $ -90.206$ & $ -92.423$  \\
  2458928.246  &   2020/03/19.746  &  0.086  & $ -89.386$ & $ -88.900$ & $ -89.468$ & $ -89.321$ & $ -89.757$ & $ -89.437$ & $ -88.908$ & $ -91.165$ & $ -89.852$ & $ -93.099$  \\
  2459029.542  &   2020/06/29.042  &  0.296  & $ -88.445$ & $ -87.144$ & $ -88.782$ & $ -88.692$ & $ -89.248$ & $ -88.873$ & $ -88.282$ & $ -90.424$ & $ -88.476$ & $ -91.352$  \\
  2459045.538  &   2020/07/15.038  &  0.329  & $ -88.550$ & $ -87.657$ & $ -88.557$ & $ -88.563$ & $ -89.015$ & $ -88.675$ & $ -88.035$ & $ -90.338$ & $ -88.990$ & $ -91.735$  \\
  2459062.461  &   2020/07/31.961  &  0.364  & $ -90.054$ & $ -89.489$ & $ -90.100$ & $ -90.086$ & $ -90.507$ & $ -90.298$ & $ -89.912$ & $ -91.618$ & $ -90.938$ & $ -93.832$  \\
  2459063.521  &   2020/08/02.021  &  0.366  & $ -90.318$ & $ -89.774$ & $ -90.125$ & $ -90.245$ & $ -90.703$ & $ -90.465$ & $ -90.134$ & $ -91.953$ & $ -90.898$ & $ -93.966$  \\
  2459067.485  &   2020/08/05.985  &  0.374  & $ -90.994$ & $ -90.127$ & $ -90.851$ & $ -90.924$ & $ -91.380$ & $ -91.123$ & $ -90.781$ & $ -92.638$ & $ -91.394$ & $ -94.456$  \\
  2459074.547  &   2020/08/13.047  &  0.389  & $ -91.182$ & $ -90.236$ & $ -91.284$ & $ -91.309$ & $ -91.668$ & $ -91.523$ & $ -90.957$ & $ -93.152$ & $ -91.586$ & $ -94.642$  \\
  2459075.506  &   2020/08/14.006  &  0.391  & $ -91.504$ & $ -90.107$ & $ -91.115$ & $ -91.353$ & $ -91.602$ & $ -91.484$ & $ -90.909$ & $ -93.114$ & $ -91.685$ & $ -94.551$  \\
  2459090.373  &   2020/08/28.873  &  0.422  & $ -92.059$ & $ -91.213$ & $ -92.117$ & $ -92.230$ & $ -92.654$ & $ -92.507$ & $ -92.079$ & $ -93.790$ & $ -92.448$ & $ -95.882$  \\
  2459096.486  &   2020/09/03.986  &  0.434  & $ -92.649$ & $ -91.666$ & $ -92.763$ & $ -92.833$ & $ -93.307$ & $ -93.048$ & $ -92.490$ & $ -94.670$ & $ -93.155$ & $ -96.547$  \\
  2459105.508  &   2020/09/13.008  &  0.453  & $ -94.196$ & $ -93.320$ & $ -94.239$ & $ -94.350$ & $ -94.713$ & $ -94.587$ & $ -94.156$ & $ -96.049$ & $ -94.893$ & $ -97.830$  \\
  2459106.415  &   2020/09/13.915  &  0.455  & $ -94.423$ & $ -93.498$ & $ -94.426$ & $ -94.542$ & $ -94.837$ & $ -94.600$ & $ -94.316$ & $ -96.309$ & $ -94.834$ & $ -97.926$  \\
  2459108.412  &   2020/09/15.912  &  0.459  & $ -94.494$ & $ -93.651$ & $ -94.636$ & $ -94.721$ & $ -95.202$ & $ -94.877$ & $ -94.421$ & $ -96.596$ & $ -95.248$ & $ -98.256$  \\
  2459146.332  &   2020/10/23.832  &  0.538  & $ -97.464$ & $ -96.620$ & $ -97.563$ & $ -97.738$ & $ -98.156$ & $ -98.225$ & $ -97.031$ & $ -99.331$ & $ -98.070$ & $-100.916$  \\
  2459150.411  &   2020/10/27.911  &  0.546  & $ -97.961$ & $ -97.297$ & $ -98.105$ & $ -98.353$ & $ -98.795$ & $ -98.531$ & $ -98.069$ & $ -99.868$ & $ -98.542$ & $-101.628$  \\
  2459154.382  &   2020/10/31.882  &  0.554  & $ -98.500$ & $ -97.675$ & $ -98.725$ & $ -98.757$ & $ -99.221$ & $ -98.934$ & $ -98.555$ & $-100.491$ & $ -98.892$ & $-101.796$  \\
  2459163.305  &   2020/11/09.805  &  0.573  & $ -98.651$ & $ -97.658$ & $ -98.984$ & $ -99.012$ & $ -99.384$ & $ -99.087$ & $ -98.533$ & $-100.715$ & $ -99.207$ & $-102.302$  \\
  2459178.245  &   2020/11/24.745  &  0.604  & $ -99.419$ & $ -99.425$ & $-100.456$ & $-100.640$ & $-101.095$ & $-100.134$ & $-100.276$ & $-102.256$ & $-100.769$ & $-103.658$  \\
  2459179.269  &   2020/11/25.769  &  0.606  & $-100.342$ & $ -99.480$ & $-100.674$ & $-100.816$ & $-101.178$ & $-100.969$ & $-100.388$ & $-102.308$ & $-100.884$ & $-103.805$  \\
  2459180.388  &   2020/11/26.888  &  0.608  & $-100.724$ & $ -99.464$ & $-100.794$ & $-100.879$ & $-101.284$ & $-101.057$ & $-100.506$ & $-102.408$ & $-100.966$ & $-104.141$  \\
  2459185.349  &   2020/12/01.849  &  0.618  & $-100.652$ & $ -99.732$ & $-100.953$ & $-101.121$ & $-101.525$ & $-101.341$ & $-100.807$ & $-102.837$ & $-101.409$ & $-104.103$  \\
  2459195.337  &   2020/12/11.837  &  0.639  & $-100.929$ & $-100.124$ & $-101.164$ & $-101.140$ & $-101.479$ & $-100.981$ & $-100.755$ & $-102.940$ & $-101.763$ & $-104.611$  \\
  2459203.355  &   2020/12/19.855  &  0.656  & $-100.965$ & $-100.147$ & $-101.205$ & $-101.314$ & $-101.615$ & $-101.314$ & $-100.902$ & $-102.956$ & $-101.655$ & $-104.816$  \\
  2459216.304  &   2021/01/01.804  &  0.683  & $-101.904$ & $-101.111$ & $-102.219$ & $-102.373$ & $-102.558$ & $-102.135$ & $-101.939$ & $-103.997$ & $-102.512$ & $-105.272$  \\
  2459224.429  &   2021/01/09.929  &  0.699  & $-101.699$ & $-100.956$ & $-101.948$ & $-102.063$ & $-102.544$ & $-102.067$ & $-101.204$ & $-103.854$ & $-102.800$ & $-105.642$  \\
  2459226.207  &   2021/01/11.707  &  0.703  & $-102.216$ & $-101.277$ & $-101.806$ & $-102.247$ & $-102.355$ & $-102.072$ & $-101.709$ & $-103.819$ & $-102.914$ & $-105.904$  \\
  2459246.276  &   2021/01/31.776  &  0.745  & $-102.520$ & $-101.513$ & $-102.721$ & $-102.976$ & $-103.239$ & $-102.844$ & $-102.397$ & $-104.721$ & $-103.191$ & $-106.213$  \\
  2459268.234  &   2021/02/22.734  &  0.790  & $-102.291$ & $-101.572$ & $-102.495$ & $-102.589$ & $-103.117$ & $-102.864$ & $-102.289$ & $-104.314$ & $-103.058$ & $-105.798$  \\
  2459271.248  &   2021/02/25.748  &  0.796  & $-102.117$ & $-100.977$ & $-101.985$ & $-102.485$ & $-102.588$ & $-102.776$ & $-101.596$ & $-104.014$ & $-102.634$ & $-105.638$  \\
  2459275.244  &   2021/03/01.744  &  0.805  & $-102.122$ & $-101.517$ & $-102.193$ & $-102.081$ & $-102.412$ & $-102.431$ & $-102.027$ & $-103.985$ & $-102.784$ & $-105.378$  \\
  2459344.565  &   2021/05/10.065  &  0.948  & $ -97.870$ & $ -97.448$ & $ -97.671$ & $ -97.590$ & $ -97.891$ & $ -97.674$ & $ -97.574$ & $ -99.327$ & $ -98.696$ & $-101.306$  \\
  2459388.500  &   2021/06/23.000  &  0.039  & $ -92.442$ & $ -91.838$ & $ -92.638$ & $ -92.516$ & $ -92.952$ & $ -92.502$ & $ -92.312$ & $ -94.170$ & $ -93.570$ & $ -95.654$  \\
  2459392.538  &   2021/06/27.038  &  0.048  & $ -91.992$ & $ -91.281$ & $ -92.048$ & $ -92.067$ & $ -92.401$ & $ -92.181$ & $ -91.726$ & $ -93.784$ & $ -92.841$ & $ -95.512$  \\
  2459620.262  &   2022/02/09.762  &  0.520  & $ -96.121$ & $ -95.234$ & $ -96.073$ & $ -95.639$ & $ -96.524$ & $ -96.050$ & $ -95.524$ & $ -97.111$ & $ -96.638$ & $ -98.029$  \\
  2459624.261  &   2022/02/13.761  &  0.528  & $ -96.217$ & $ -95.364$ & $ -96.307$ & $ -96.532$ & $ -96.819$ & $ -96.632$ & $ -96.098$ & $ -97.353$ & $ -96.783$ & $ -99.876$  \\
  2459650.236  &   2022/03/11.736  &  0.582  & $ -98.709$ & $ -98.505$ & $ -99.005$ & $ -99.228$ & $ -99.298$ & $ -99.403$ & $ -98.704$ & $ -99.937$ & $ -99.639$ & $-102.261$  \\
  2459698.566  &   2022/04/29.066  &  0.682  & $-101.800$ & $-101.271$ & $-101.634$ & $-100.861$ & $-102.399$ & $-101.919$ & $-101.503$ & $-102.754$ & $-102.129$ & $-104.318$  \\
  2459952.259  &   2023/01/07.759  &  0.208  & $ -87.028$ & $ -85.973$ & $ -87.218$ & $ -87.113$ & $ -87.509$ & $ -87.136$ & $ -86.707$ & $ -88.798$ & $ -87.256$ & $ -89.911$  \\ 
\noalign{\smallskip}
\hline
\end{tabular}
\begin{flushleft}
{\bf Notes.}\\ 
HJD stands for the heliocentric
Julian date, 'dateobs' is the civil date [UT] in the standard
format, yyyy/mm/dd.ddd, and 'phase' is the corresponding orbital
phase according to Eq.~(\ref{eq:ephem}).
\end{flushleft}
\end{table*}

\begin{table*}[t!]
\small
\setlength{\tabcolsep}{5pt}
   \caption{
Absorbed fluxes [$10^{-12}$\ecs] of ten Fe\,I absorption lines from our 53 spectra (Sect.~\ref{ss:orbvar}).
           }
\label{tabAp:fluxes}
\centering
\begin{tabular}{ccccccccccccc}
\hline
\hline
\noalign{\smallskip}
 HJD      &     dateobs      &  phase  &  5151\,\AA  &    5340\,\AA &    5371\,\AA &    5406\,\AA &    5430\,\AA &    5435\,\AA &    5501\,\AA &    5507\,\AA &    6412\,\AA &   6469\,\AA \\
\noalign{\smallskip}
\hline
\noalign{\smallskip}
  2457743.439  &   2016/12/20.939  &  0.630  & $ 2.136$ & $ 1.834$ & $ 3.512$ & $ 3.036$ & $ 3.672$ & $ 3.032$ & $ 3.079$ & $ 3.191$ & $ 1.238$ & $ 0.650$  \\
  2457914.508  &   2017/06/10.008  &  0.985  & $ 1.371$ & $ 0.945$ & $ 2.780$ & $ 2.437$ & $ 2.663$ & $ 2.074$ & $ 2.099$ & $ 2.499$ & $ 1.199$ & $ 0.829$  \\
  2457924.512  &   2017/06/20.012  &  0.006  & $ 1.280$ & $ 0.847$ & $ 2.656$ & $ 2.275$ & $ 2.494$ & $ 1.965$ & $ 1.974$ & $ 2.228$ & $ 1.068$ & $ 0.753$  \\
  2457926.523  &   2017/06/22.023  &  0.010  & $ 1.341$ & $ 0.888$ & $ 2.705$ & $ 2.422$ & $ 2.634$ & $ 2.085$ & $ 2.074$ & $ 2.438$ & $ 1.174$ & $ 0.812$  \\
  2457941.517  &   2017/07/07.017  &  0.041  & $ 1.476$ & $ 0.863$ & $ 2.881$ & $ 2.627$ & $ 2.844$ & $ 2.176$ & $ 2.150$ & $ 2.508$ & $ 1.109$ & $ 0.761$  \\
  2457964.549  &   2017/07/30.049  &  0.089  & $ 1.374$ & $ 0.866$ & $ 2.821$ & $ 2.474$ & $ 2.716$ & $ 2.136$ & $ 2.081$ & $ 2.408$ & $ 1.142$ & $ 0.827$  \\
  2458073.421  &   2017/11/15.921  &  0.314  & $ 1.686$ & $ 1.079$ & $ 3.340$ & $ 2.919$ & $ 3.268$ & $ 2.474$ & $ 2.314$ & $ 2.790$ & $ 1.348$ & $ 0.926$  \\
  2458080.387  &   2017/11/22.887  &  0.329  & $ 1.649$ & $ 1.120$ & $ 3.278$ & $ 2.805$ & $ 3.185$ & $ 2.435$ & $ 2.384$ & $ 2.806$ & $ 1.293$ & $ 0.978$  \\
  2458370.558  &   2018/09/09.058  &  0.930  & $ 1.441$ & $ 0.954$ & $ 2.979$ & $ 2.548$ & $ 2.839$ & $ 2.224$ & $ 2.115$ & $ 2.531$ & $ 1.211$ & $ 0.720$  \\
  2458900.210  &   2020/02/20.710  &  0.028  & $ 1.430$ & $ 1.012$ & $ 2.939$ & $ 2.525$ & $ 2.797$ & $ 2.125$ & $ 2.060$ & $ 2.431$ & $ 0.894$ & $ 0.524$  \\
  2458911.235  &   2020/03/02.735  &  0.050  & $ 1.330$ & $ 0.856$ & $ 2.668$ & $ 2.310$ & $ 2.638$ & $ 2.013$ & $ 1.861$ & $ 2.148$ & $ 0.957$ & $ 0.610$  \\
  2458914.242  &   2020/03/05.742  &  0.057  & $ 1.229$ & $ 0.719$ & $ 2.458$ & $ 2.132$ & $ 2.490$ & $ 1.903$ & $ 1.803$ & $ 2.044$ & $ 1.016$ & $ 0.546$  \\
  2458918.238  &   2020/03/09.738  &  0.065  & $ 1.207$ & $ 0.788$ & $ 2.510$ & $ 2.116$ & $ 2.441$ & $ 1.795$ & $ 1.800$ & $ 2.040$ & $ 0.861$ & $ 0.482$  \\
  2458926.248  &   2020/03/17.748  &  0.082  & $ 1.212$ & $ 0.749$ & $ 2.542$ & $ 2.142$ & $ 2.443$ & $ 1.853$ & $ 1.822$ & $ 2.222$ & $ 0.865$ & $ 0.432$  \\
  2458927.243  &   2020/03/18.743  &  0.084  & $ 1.180$ & $ 0.762$ & $ 2.469$ & $ 2.119$ & $ 2.396$ & $ 1.819$ & $ 1.810$ & $ 2.078$ & $ 0.881$ & $ 0.559$  \\
  2458928.246  &   2020/03/19.746  &  0.086  & $ 1.186$ & $ 0.715$ & $ 2.471$ & $ 2.118$ & $ 2.380$ & $ 1.819$ & $ 1.782$ & $ 2.068$ & $ 1.011$ & $ 0.604$  \\
  2459029.542  &   2020/06/29.042  &  0.296  & $ 1.571$ & $ 1.065$ & $ 3.035$ & $ 2.599$ & $ 2.901$ & $ 2.249$ & $ 2.069$ & $ 2.438$ & $ 1.024$ & $ 0.672$  \\
  2459045.538  &   2020/07/15.038  &  0.329  & $ 1.449$ & $ 0.971$ & $ 2.821$ & $ 2.435$ & $ 2.711$ & $ 2.090$ & $ 1.992$ & $ 2.337$ & $ 1.047$ & $ 0.659$  \\
  2459062.461  &   2020/07/31.961  &  0.364  & $ 1.385$ & $ 0.950$ & $ 2.669$ & $ 2.358$ & $ 2.665$ & $ 2.018$ & $ 1.905$ & $ 2.378$ & $ 1.009$ & $ 0.576$  \\
  2459063.521  &   2020/08/02.021  &  0.366  & $ 1.390$ & $ 0.985$ & $ 2.724$ & $ 2.375$ & $ 2.674$ & $ 2.015$ & $ 1.934$ & $ 2.407$ & $ 0.963$ & $ 0.598$  \\
  2459067.485  &   2020/08/05.985  &  0.374  & $ 1.375$ & $ 0.970$ & $ 2.744$ & $ 2.396$ & $ 2.681$ & $ 2.032$ & $ 1.977$ & $ 2.364$ & $ 0.926$ & $ 0.643$  \\
  2459074.547  &   2020/08/13.047  &  0.389  & $ 1.402$ & $ 1.013$ & $ 2.767$ & $ 2.417$ & $ 2.716$ & $ 2.067$ & $ 2.023$ & $ 2.465$ & $ 0.940$ & $ 0.587$  \\
  2459075.506  &   2020/08/14.006  &  0.391  & $ 1.556$ & $ 0.977$ & $ 2.798$ & $ 2.423$ & $ 2.675$ & $ 2.065$ & $ 2.014$ & $ 2.389$ & $ 0.925$ & $ 0.568$  \\
  2459090.373  &   2020/08/28.873  &  0.422  & $ 1.730$ & $ 1.326$ & $ 3.071$ & $ 2.687$ & $ 2.958$ & $ 2.299$ & $ 2.139$ & $ 2.639$ & $ 0.949$ & $ 0.646$  \\
  2459096.486  &   2020/09/03.986  &  0.434  & $ 1.735$ & $ 1.377$ & $ 3.180$ & $ 2.792$ & $ 3.113$ & $ 2.404$ & $ 2.276$ & $ 2.784$ & $ 0.767$ & $ 0.408$  \\
  2459105.508  &   2020/09/13.008  &  0.453  & $ 1.950$ & $ 1.410$ & $ 3.357$ & $ 2.926$ & $ 3.253$ & $ 2.516$ & $ 2.351$ & $ 2.882$ & $ 0.964$ & $ 0.610$  \\
  2459106.415  &   2020/09/13.915  &  0.455  & $ 1.962$ & $ 1.424$ & $ 3.452$ & $ 2.934$ & $ 3.260$ & $ 2.515$ & $ 2.374$ & $ 2.797$ & $ 0.945$ & $ 0.655$  \\
  2459108.412  &   2020/09/15.912  &  0.459  & $ 1.978$ & $ 1.587$ & $ 3.485$ & $ 3.097$ & $ 3.429$ & $ 2.636$ & $ 2.496$ & $ 3.013$ & $ 0.857$ & $ 0.496$  \\
  2459146.332  &   2020/10/23.832  &  0.538  & $ 1.785$ & $ 1.319$ & $ 3.067$ & $ 2.682$ & $ 3.096$ & $ 2.369$ & $ 2.139$ & $ 2.696$ & $ 1.080$ & $ 0.669$  \\
  2459150.411  &   2020/10/27.911  &  0.546  & $ 1.765$ & $ 1.266$ & $ 3.033$ & $ 2.644$ & $ 2.888$ & $ 2.199$ & $ 2.139$ & $ 2.504$ & $ 1.002$ & $ 0.632$  \\
  2459154.382  &   2020/10/31.882  &  0.554  & $ 1.805$ & $ 1.382$ & $ 3.120$ & $ 2.749$ & $ 2.968$ & $ 2.370$ & $ 2.257$ & $ 2.658$ & $ 1.074$ & $ 0.641$  \\
  2459163.305  &   2020/11/09.805  &  0.573  & $ 1.708$ & $ 1.375$ & $ 2.946$ & $ 2.627$ & $ 2.823$ & $ 2.203$ & $ 2.165$ & $ 2.561$ & $ 1.021$ & $ 0.662$  \\
  2459178.245  &   2020/11/24.745  &  0.604  & $ 2.721$ & $ 1.733$ & $ 3.909$ & $ 3.285$ & $ 3.589$ & $ 3.067$ & $ 2.615$ & $ 3.095$ & $ 1.069$ & $ 0.627$  \\
  2459179.269  &   2020/11/25.769  &  0.606  & $ 2.259$ & $ 1.795$ & $ 4.029$ & $ 3.503$ & $ 3.852$ & $ 3.208$ & $ 2.897$ & $ 3.381$ & $ 0.935$ & $ 0.519$  \\
  2459180.388  &   2020/11/26.888  &  0.608  & $ 2.213$ & $ 1.739$ & $ 3.930$ & $ 3.380$ & $ 3.630$ & $ 2.835$ & $ 2.624$ & $ 3.156$ & $ 1.047$ & $ 0.663$  \\
  2459185.349  &   2020/12/01.849  &  0.618  & $ 2.097$ & $ 1.558$ & $ 3.464$ & $ 2.984$ & $ 3.262$ & $ 2.645$ & $ 2.437$ & $ 2.806$ & $ 1.106$ & $ 0.677$  \\
  2459195.337  &   2020/12/11.837  &  0.639  & $ 1.803$ & $ 1.354$ & $ 3.197$ & $ 2.796$ & $ 3.038$ & $ 2.406$ & $ 2.257$ & $ 2.685$ & $ 1.064$ & $ 0.699$  \\
  2459203.355  &   2020/12/19.855  &  0.656  & $ 1.949$ & $ 1.569$ & $ 3.368$ & $ 3.008$ & $ 3.248$ & $ 2.520$ & $ 2.423$ & $ 2.830$ & $ 1.063$ & $ 0.717$  \\
  2459216.304  &   2021/01/01.804  &  0.683  & $ 1.963$ & $ 1.348$ & $ 3.518$ & $ 2.964$ & $ 3.250$ & $ 2.561$ & $ 2.471$ & $ 2.794$ & $ 1.031$ & $ 0.680$  \\
  2459224.429  &   2021/01/09.929  &  0.699  & $ 1.984$ & $ 1.438$ & $ 3.540$ & $ 2.965$ & $ 3.308$ & $ 2.632$ & $ 2.498$ & $ 2.874$ & $ 1.027$ & $ 0.681$  \\
  2459226.207  &   2021/01/11.707  &  0.703  & $ 1.935$ & $ 1.309$ & $ 3.436$ & $ 2.930$ & $ 3.160$ & $ 2.478$ & $ 2.424$ & $ 2.710$ & $ 1.137$ & $ 0.700$  \\
  2459246.276  &   2021/01/31.776  &  0.745  & $ 1.623$ & $ 1.200$ & $ 3.375$ & $ 2.866$ & $ 3.162$ & $ 2.495$ & $ 2.464$ & $ 2.785$ & $ 1.157$ & $ 0.754$  \\
  2459268.234  &   2021/02/22.734  &  0.790  & $ 1.591$ & $ 1.076$ & $ 3.169$ & $ 2.747$ & $ 3.035$ & $ 2.398$ & $ 2.241$ & $ 2.628$ & $ 1.052$ & $ 0.748$  \\
  2459271.248  &   2021/02/25.748  &  0.796  & $ 1.587$ & $ 1.117$ & $ 3.102$ & $ 2.697$ & $ 3.301$ & $ 2.463$ & $ 2.356$ & $ 2.642$ & $ 1.239$ & $ 0.769$  \\
  2459275.244  &   2021/03/01.744  &  0.805  & $ 1.564$ & $ 1.083$ & $ 3.140$ & $ 2.719$ & $ 3.076$ & $ 2.430$ & $ 2.300$ & $ 2.738$ & $ 1.206$ & $ 0.709$  \\
  2459344.565  &   2021/05/10.065  &  0.948  & $ 1.503$ & $ 1.073$ & $ 3.108$ & $ 2.773$ & $ 3.102$ & $ 2.409$ & $ 2.311$ & $ 2.643$ & $ 1.215$ & $ 0.732$  \\
  2459388.500  &   2021/06/23.000  &  0.039  & $ 1.467$ & $ 1.018$ & $ 3.077$ & $ 2.692$ & $ 3.006$ & $ 2.394$ & $ 2.250$ & $ 2.603$ & $ 1.373$ & $ 0.756$  \\
  2459392.538  &   2021/06/27.038  &  0.048  & $ 1.402$ & $ 0.903$ & $ 2.962$ & $ 2.581$ & $ 2.856$ & $ 2.258$ & $ 2.168$ & $ 2.595$ & $ 1.420$ & $ 0.797$  \\
  2459620.262  &   2022/02/09.762  &  0.520  & $ 1.326$ & $ 1.119$ & $ 2.983$ & $ 2.423$ & $ 2.880$ & $ 2.184$ & $ 2.140$ & $ 2.156$ & $ 1.058$ & $ 0.502$  \\
  2459624.261  &   2022/02/13.761  &  0.528  & $ 1.375$ & $ 1.076$ & $ 2.876$ & $ 2.472$ & $ 2.837$ & $ 2.192$ & $ 2.172$ & $ 2.142$ & $ 1.123$ & $ 0.615$  \\
  2459650.236  &   2022/03/11.736  &  0.582  & $ 1.450$ & $ 1.035$ & $ 2.891$ & $ 2.578$ & $ 2.995$ & $ 2.340$ & $ 2.311$ & $ 2.199$ & $ 1.029$ & $ 0.609$  \\
  2459698.566  &   2022/04/29.066  &  0.682  & $ 1.248$ & $ 0.871$ & $ 2.717$ & $ 2.562$ & $ 2.617$ & $ 1.994$ & $ 1.931$ & $ 1.981$ & $ 1.343$ & $ 0.747$  \\
  2459952.259  &   2023/01/07.759  &  0.208  & $ 1.698$ & $ 1.349$ & $ 3.490$ & $ 3.083$ & $ 3.458$ & $ 2.702$ & $ 2.552$ & $ 3.058$ & $ 1.073$ & $ 0.880$  \\  
\noalign{\smallskip}
\hline
\end{tabular}
\end{table*}

\end{appendix}


\end{document}